\documentclass[letterpaper, twocolumn, pre, floatfix]{revtex4-1}
\usepackage{amssymb}
\usepackage{amsmath}
\usepackage{bm}
\usepackage{graphics}
\usepackage{graphicx}
\usepackage{psfrag}
\usepackage{subfigure}

\newcommand{\eqn}[0]{Eq.}
\newcommand{\fig}[0]{Fig.}
\newcommand{\scn}[0]{Sec.}

\newcommand{\eqnref}[1]{\eqn~\eqref{#1}}
\newcommand{\figref}[1]{\fig~\ref{#1}}
\newcommand{\scnref}[1]{\scn~\ref{#1}}

\newcommand{\sqr}{\mathop{\mathrm{Sqr}}}
\newcommand{\sign}{\mathop{\mathrm{signum}}}
\newcommand{\abs}[1]{\left|#1\right|}

\newcommand{\arccot}[1]{\mathop{\mathrm{arccot}\left(#1\right)}}
\newcommand{\arctanh}[1]{\mathop{\mathrm{arctanh}\left(#1\right)}}

\newcommand{\cvec}[1]{\begin{pmatrix} #1 \end{pmatrix}}

\renewcommand{\vec}[1]{\mathbf{#1}}
\let\oldhat\hat
\renewcommand{\hat}[1]{\oldhat{\mathbf{#1}}}

\begin{document}
\title{Survey of Morphologies Formed in the Wake of an Enslaved Phase-Separation Front in Two Dimensions}
\author{E.~M.~Foard and A.~J.~Wagner}
\affiliation{Department of Physics, North Dakota State University, Fargo, ND 58105}

\begin{abstract}
A phase-separation front will leave in its wake a phase-separated morphology that differs markedly from homogeneous phase-separation morphologies. For a purely diffusive system such a front, moving with constant velocity, will generate very regular, non-equilibrium structures. We present here a numerical study of these fronts using a lattice Boltzmann method. In two dimensions these structures are regular stripes or droplet arrays. In general the kind and orientation of the selected morphology and the size of the domains depends on the speed of the front as well as the composition of the material overtaken by the phase-separation front. We present a survey of morphologies as a function of these two parameters. We show that the resulting morphologies are initial condition dependent. We then examine which of the potential morphologies is the most stable. An analytical analysis for symmetrical compositions predicts the transition point from orthogonal to parallel stripes.
\end{abstract}

\maketitle

\section{Introduction}\label{scn:introduction}
Phase-separation is a ubiquitous phenomenon observed in a wide variety of systems. The theoretical analysis of phase-separation has mostly focused on the case where the system is homogeneously quenched, i.e. moved instantaneously and uniformly from a mixed state to a state where the system will separate into different phases. A good overview of this theoretical work is provided in the book by Onuki\cite{onuki-2002}.
 
In many practically occurring systems phase-separation does not occur everywhere at once, but rather starts in a specific region and from this place successively invades the system. We refer to the surface of transition between the mixed and the separated regions as the phase-separation front. The resulting morphologies formed in the wake of a phase-separation front can differ significantly from the structures resulting from homogeneous phase-separation.

Our interest in phase-separation fronts arose from an investigation of immersion precipitation membranes\cite{akthakul-2005}. In these systems a polymer-solvent mixture, applied thinly to a substrate, is immersed in water. As solvent leaks into the water and as water enters the polymer-solvent mixture phase-separation is induced.  It starts from the water--polymer/solvent interface until all the solvent migrated into the water bath and a porous, asymmetric, polymer structure is formed. This structure shows a thin initial layer of polymer on the surface that will, ideally, show small holes. Below this layer one typically finds much larger structures. To understand why such structures are formed Akthakul et al. simulated immersion precipitation membrane formation in a lattice Boltzmann framework\cite{akthakul-2005}.  Shortly thereafter Zhou and Powell examined the same system using a finite difference approach\cite{zhou-2006}.  More recently Wang et al. used a dissipative particle dynamics method to simulate a the effects of varying polymer chain length on the formation of immersion precipitation membranes\cite{wang-2007}.  However, the system proved much too complicated to allow the simulations to generate significant insight into the main phenomena governing the membrane formation.

The simulations by Akthakul et al. suggested to us the possibility that the main factor controlling the structure formation was the dynamics of the phase-separation front. However, we found that the dynamics of phase-separation fronts in polymer systems was poorly studied. This inspired us to investigate the simplest possible model for a phase-separation front, i.e. a phase-separation front induced by sharp front of a control parameter (solvent concentration in the immersion precipitation example) moving with a prescribed speed.

Phase-separation fronts are also of paramount importance in many eutectic alloys, where the physics of the front is responsible for the formation and orientation of ordered structures, and phase-separation goes hand in hand with a solidification problem.  Jackson and Hunt analyzed the formation of the lamellar (essentially two-dimensional) and rod structures sometimes formed in eutectics.  They modeled the dynamics as a steady-state diffusion driven phase-separation which occurs directly ahead of the solidification front\cite{jackson-1966}.  They confirmed the earlier observed relation between the solidification front speed and the lamellar spacing, and in doing so refined some of the earlier theoretical findings.  Much of the work on eutectic solidification fronts following Jackson and Hunt involved adding refinements to their model, increasing complexity to make it more like the real alloys under investigation.  For instance, the inclusion of a convection layer just ahead of the solidification front by Verhoeven and Homer\cite{verhoeven-1970}.

Other researchers took an opposite approach; developing simple models that could be simulated with numerical methods.  Early work by Ball and Essery simulating front-induced phase-separation of a binary mixture noted the remarkable difference between phase-separation structures formed by fronts and those formed by homogeneous phase separation\cite{ball-1990}.  Their model consists of an underlying Ginzburg-Landau free energy, similar to our model in this paper.  However, their control parameter is designed to mimic a heat-diffusion process, in analogy to the temperature front in eutectic solidification.  For slow thermal diffusion they observed the lamellar structure familiar to eutecticts researchers, in addition to a lamellar structure oriented orthogonally to the motion of the temperature front when thermal diffusion was fast.

More recently, realistic phase separation fronts induced by a control-parameter with its own dynamics, have also been studied by Gonnella at al\cite{gonnella-2008-41, gonnella-2010}. They examined a binary fluid where a phase-separation is induced by temperature change at the walls. The temperature then diffuses into the finite system, inducing a phase-separation front. In this systems the shape and speed of the control parameter (in this case the temperature) are space and time dependent, complicating the analysis of the resulting structures.

Another example of a realistic modeling of a phase-separation front is given by K\"opf \emph{et al.} who examined pattern formation in monolayer transfer for systems with substrate mediated condensation\cite{kopf-2010}.  Here the similar patterns to the ones predicted in this paper are observed in a more complicated system were additional hydrodynamic effects lead to a condensation of surfactants in the deposition layer.  In this case the surfactant concentration in the deposited later differs substantially from the lipid concentration in the free film so that the concentration is now dependent on the speed with which the phase-separation front advances.

Hantz and Bir\'o developed a further simplified model of a phase-separation front as a moving Gaussian source in a diffusive system\cite{hantz-2006}.  By decoupling phase separation from a dynamic control parameter, they are able to control the phase separation front more directly.  In particular, Hantz and Bir\'o assembled a rotating front, where the front speed is a function of distance from the axis, and the direction of motion of the front changes as it sweeps across the material.  They found, in addition to the expected perpendicular and parallel lamella structures at the respective slow and fast front speeds, that the lamellar structures could bend to continue growing perpendicular to the front.  They also observed that the parallel lamella formed with width and spacing dependent on the speed of the front, and they noted that this was consistent with experimental observations of Liesegang patterns.  Liesegang patterns are highly ordered structures formed in the wake of an electrolyte reaction front in a gel\cite{liesegang-1896}.  Antal \emph{et al.} had earlier used a similar Gaussian source front to produce patterns with position and spacing laws consistent with Liesegang patterns\cite{PhysRevLett.83.2880}.  In an earlier paper of ours, we prove analytically that Liesegang patterns are reproducible by a similar model\cite{foard-2010}.  Refer to the citations contained in that earlier paper for other models which have been demonstrated to produce Liesegang patterns under the proper conditions.

An even simpler model of a moving phase separation front was used by Furukawa for simulating the formation of phase separation induced morphologies\cite{furukawa-1990}.   The front in his model is an abrupt change in the control parameter moving at a constant average speed.  Similar to Ball and Essery, Furukawa used a Ginzburg-Landau free energy.  The model by Furukawa is very similar to the one used here, with a few minor differences.  The implementation of our model in numerical simulation, however, is quite different as we will show in \scnref{scn:simulation}.

In this paper we focus on the simplest possible case: we consider a purely diffusive system, as hydrodynamics adds additional complexity to this problem. We consider fronts moving with a constant speed, since this allows us to separate transient phenomena from generic phenomena, simply by observing the front after a sufficiently long time. Furthermore we consider a sharp front, so we do not have to consider the effects of the shape of the front. For this paper we will focus on two-dimensional systems. Despite its simplicity such fronts still exhibit a rich collection of behaviors, as we will show in this paper.

This work is an extension of our work on phase separation fronts in one-dimensions. In the one dimensional case phase-separation fronts will leave in their wake alternating domains, and the only remaining question is the size of these domains\cite{foard-2010,foard-2009}. We were able to show that this problem could be solved analytically, at least in the limit of small velocities. For two-dimensional systems one possible solution are stripes oriented in parallel with the front, which are essentially the same as the one-dimensional systems observed earlier.

\section{Model}\label{scn:model}
The model we use is essentially the same as the one we presented in our earlier paper\cite{foard-2009} extended to two dimensions. We therefore give the most important aspects in short here:

To construct a model for phase separation fronts in two dimensions we consider a mixture of two materials, an $\mathcal{A}$-type and a $\mathcal{B}$-type, in an incompressible mixture such that the total density $\rho = \rho_\mathcal{A}(\vec{r},t) + \rho_\mathcal{B}(\vec{r},t)$ is a constant.  In this paper the position vector $\vec{r}=(x,y)$ is two-dimensional.  The order parameter for this system is the relative concentration:
\begin{equation}
	\phi(\vec{r},t) = \frac{\rho_\mathcal{A}(\vec{r},t) - \rho_\mathcal{B}(\vec{r},t)}{\rho} \,.
\end{equation}
For simplicity of the model, we choose a $\phi^4$-type mixing free energy\cite{ginzburg-1950}:
\begin{eqnarray}
	\label{eqn:free-energy}
	\nonumber F=\int\!\!dx && \left[ \frac{a(\vec{r},t)}{2}\phi(\vec{r},t)^2 + \frac{b(\vec{r},t)}{4}\phi(\vec{r},t)^4 + \right. \\
	&& \left. c(\vec{r},t)\phi(\vec{r},t) + \frac{\kappa(\vec{r},t)}{2}(\nabla\phi(\vec{r},t))^2 \right] \,.
\end{eqnarray}
The $c$ term, linear in the order parameter, adds a constant to the chemical potential for spatially homogeneous systems.  However, in the equation of motion only gradients of the chemical potential enter the dynamics, so that a constant added to the chemical potential does not alter the dynamics of the order parameter.  It is included here, however, since we will consider different values of $c$ across the front, which does influence the dynamics.

A phase separation front constitutes a spatio-temporal change in the control parameter $a(\vec{r},t)$ such that the free energy at a given location transitions from a mixing state ($a>0$) with a single minimum, to a separating state ($a<0$) with two minima.  Again for simplicity, we choose the transition to be an abrupt jump from a single positive mixing value $a_M>0$ to a single negative separating value $a_S<0$.  The transition moves with constant velocity $\vec{u}$, and is flat perpendicular to $\vec{u}$.  The other control parameters are similarly two-valued, with an abrupt transition at the front:
\begin{eqnarray}
\label{eqn:parameter-a}
	\nonumber a(\vec{r},t) &=& a_S + (a_M - a_S) \Theta\left[(\vec{r} + \vec{r}_0 + \vec{u}t)\cdot\hat{u} \right] \,,\\
\label{eqn:parameter-b}
	\nonumber b(\vec{r},t) &=& b_S + (b_M - b_S) \Theta\left[(\vec{r} + \vec{r}_0 + \vec{u}t)\cdot\hat{u} \right] \,,\\
\label{eqn:parameter-c}
	\nonumber c(\vec{r},t) &=& c_S + (c_M - c_S) \Theta\left[(\vec{r} + \vec{r}_0 + \vec{u}t)\cdot\hat{u} \right] \,,\\
\label{eqn:parameter-k}
	\kappa(\vec{r},t) &=& \kappa_S + (\kappa_M - \kappa_S) \Theta\left[(\vec{r} + \vec{r}_0 + \vec{u}t)\cdot\hat{u} \right] \,.
\end{eqnarray}
The \emph{mixing} and \emph{separating} values are denoted by subscripts $M$ and $S$ respectively, and $\Theta$ is the Heaviside step function.

For our diffusive system the equation of motion is
\begin{equation}
	\label{eqn:dynamics}
	\partial_t \phi(\vec{r},t)  = \nabla \cdot \left[ m(\vec{r},t) \nabla \mu(\vec{r},t) \right]\,,
\end{equation}
where $m$ is the diffusive mobility and $\mu$ is the chemical potential.
The chemical potential is derived from the free energy:
\begin{eqnarray}
	\label{eqn:chem}
	\nonumber \mu(\vec{r},t)=\frac{\delta F}{\delta \phi} &=&
	a(\vec{r},t) \phi(\vec{r},t) + b(\vec{r},t)\phi(\vec{r},t)^3 \\
	&&+ c(\vec{r},t) - \kappa(\vec{r},t) \nabla^2\phi(\vec{r},t) \,.
\end{eqnarray}
The equation of motion only considers gradients of the full chemical potential.  Since $\mu$ is itself continuous\cite[Fig. 1(b)]{foard-2009}, we need not be concerned with the computational messiness of delta functions which  result from gradients of the Heaviside function present in the parameters of \eqnref{eqn:parameter-k}.
For this model the diffusive mobility can take different values across the front:
\begin{equation}
	m(\vec{r},t) = m_S + (m_M - m_S) \Theta\left[(\vec{r} + \vec{r}_0 + \vec{u}t)\cdot\hat{u} \right] \,.
\end{equation}
For the remainder of this paper, space and time dependence will not be written explicitly, except where needed to avoid ambiguity.

Since we intend to simulate our system with a numerical method, we will only be able to examine a finite system. The way the system is described above, this would limit our analysis to times $t=l_x/u$, where $l_x$ is the length of the simulation in the direction of travel of the front.  This was a limitation of an earlier, similar model by Furukawa\cite{furukawa-1990}.  In turn, this would make it costly to investigate the system for large times. To effectively look at large times we employ a transformation here (as we did in our earlier work \cite{foard-2009}), where we keep the position of the front stationary in our simulation domain and move the sample with a constant speed $u$.  This transformation changes the diffusive equation of motion \eqnref{eqn:dynamics} to a drift-diffusion equation of motion:
\begin{equation}
	\label{eqn:simdynamics}
	\partial_t \phi(\vec{r},t) = \nabla \cdot \left[ -\phi(\vec{r},t)\vec{u} + m(\vec{r}) \nabla \mu(\vec{r},t) \right] \,.
\end{equation}
Mathematically these two approaches are entirely equivalent, but for simulation purposes the latter approach has the great advantage of allowing us to examine the front for long times.

We rewrite this model in a dimensionless form by considering the length, time, and concentration scales of spinodal decomposition for a symmetrical system ($\phi_{in} = 0$) \cite{foard-2009}.  The spinodal length is the wavelength of the concentration fluctuation that is most unstable, i.e. phase-separating the most rapidly after homogeneous phase separation is induced by the quenching a material into the spinodal region of the phase diagram.  The spinodal time is the inverse of the growth rate of those spinodal wavelength fluctuations.  These are, respectively:
\begin{equation}
	\lambda_{sp}=2\pi\sqrt{\frac{2\kappa_S}{-a_S}} \,,\,
	t_{sp}=\frac{4\kappa_S}{m_S a_S^2} \,,\,
	\phi_{eq} = \sqrt{\frac{-a_S}{b_S}} \,.
\end{equation}
One of the benefits of this non-dimensionalization is a reduction in the number of free parameters of this model to the seven following non-dimensional quantities:
\begin{eqnarray}
	\nonumber &&
	A = -\frac{a_M}{a_S} \,,\,
	B = \frac{b_M}{b_S} \,,\,
	C = \frac{c_M-c_S}{a_S\phi_{eq}} \,,\\
	&&
	M = \frac{m_M}{m_S} \,,\,
	K = \frac{\kappa_M}{\kappa_S} \,,\,
	\Phi_{in} = \frac{\phi_{in}}{\phi_{eq}} \,,\,
	U = u\frac{t_{sp}}{\lambda_{sp}} \,.
\end{eqnarray}

The non-dimensional equation of motion then becomes:
\begin{eqnarray}
	\label{eqn:nondim}
	&& \partial_T \Phi + \nabla_{\!\vec{R}} \cdot \left( \Phi \vec{U} \right) \\
	\nonumber && = \frac{1}{2\pi^2} \nabla_{\!\vec{R}} \cdot \mathcal{M} \nabla_{\!\vec{R}} \left( \mathcal{A} \Phi + \mathcal{B} \Phi^3 + \mathcal{C} - \frac{\mathcal{K}}{8\pi^2} \nabla^2_{\!\vec{R}} \Phi \right)\;,
\end{eqnarray}
where $\vec{R}=\vec{r}/\lambda_{sp}$ and $T=t/t_{sp}$ are the discrete non-dimensionalized length and time coordinates.  The capital script letters are spatially dependent functions of the non-dimensional parameters:
\begin{eqnarray}
	\mathcal{\{A,B,C,K,M\}} =
	\begin{cases}
		\{ -1,1,0,1,1 \} \,, & X < X_f \\
		\{  A,B,C,K,M \} \,, & X > X_f \\
	\end{cases}.{}\;\;\;
\end{eqnarray}
The parameters $A,B,M,K$ are chosen as $A=M=K=1$ and $B=0$.
The choice of $B=0$ is unconventional.  Its justification is as follows: Since $\phi\approx\phi_{in}$ in the mixing region, we can re-expand the free energy around $\phi=\phi_{in}$, retaining only terms to second order.  This reduces the number of free parameters.  As in our previous paper\cite{foard-2009}, we will restrict ourselves here to the case where $\mu_S(\pm\phi_{eq})=\mu_M(\phi_{in})$, so that there are no long-range diffusion dynamics ahead of the front.  Relaxing this restriction will alter the effective $\phi_{in}$ at the front and the details of the domain switching.  Our choice corresponds to $C=-A\Phi_{in}$.

This leaves as the only remaining free parameters: the initial concentration of the mixed material $\Phi_{in}$, and the speed of the advancing front $U$.  Details of the effect of changing some of the other non-dimensional parameters can be found in our previous work\cite{foard-2009}.

\section{Simulation method}\label{scn:simulation}
To simulate the drift diffusion equation (\ref{eqn:simdynamics}) we use a lattice Boltzmann (LB) method, mostly because we intend to extend our analysis to hydrodynamic systems where lattice Boltzmann methods have been shown to perform particularly well.  Lattice Boltzmann uses a discretized form of the Boltzmann transport equation \cite{qian-1992}:
\begin{equation}
	\label{eqn:LBE}
	f_i(\vec{r}+\vec{v}_i, t+1 ) - f_i(\vec{r}, t) = \frac{1}{\tau(\vec{r},t)}\left[ f_i^0 - f_i(\vec{r}, t)\right] \,.
\end{equation}
Time advances in discrete steps ($\Delta t = 1$ is implied above), and space is divided into regular cells which tile the simulation space.  The density distribution functions for individual particles are replaced by a discrete set of distribution functions $f_i(\vec{r},t)$ that represent the density of particles at position $\vec{r}$ and time $t$ moving with velocity $\vec{v}_i$.  The velocity vectors are restricted such that from any given lattice site $\vec{r}$, the transformation $\vec{r}\rightarrow\vec{r}+\vec{v}_i$, for every index $i$, always results in a $\vec{r}$ which lies on a lattice site or a boundary site.  Following C programming language conventions, the lattice sites are numbered from $0$ to $l_x-1$ in the $x$ direction.

The zero-order velocity moment of the non-equilibrium distribution functions is the order parameter:
\begin{equation}
	\sum_i f_i(\vec{r},t) = \phi(\vec{r},t) \,.
\end{equation}
The choice of the equilibrium moment distributions determines the equation of motion to be simulated.  For a drift-diffusion equation, we chose:
\begin{eqnarray}
\label{eqn:moments}
\nonumber\sum_i f_i^0(\vec{r},t) &=& \phi(\vec{r},t) \,, \\
\nonumber\sum_i f_i^0(\vec{r},t) v_{i\alpha} &=& s u_\alpha \phi(\vec{r},t) \,, \\
\sum_i f_i^0(\vec{r},t) v_{i\alpha} v_{i\beta} &=& s \mu(\vec{r},t) + s^2 u_\alpha u_\beta \phi(\vec{r},t) \delta_{\alpha\beta} \,.
\end{eqnarray}
The subscripts $\alpha$ and $\beta$ are indices for the spatial dimensions $x$ and $y$; for instance $u_\alpha$ represents the magnitude of the vector $\vec{u}$ in the $\alpha$ direction.  The time-scaling parameter $s$ introduced here could easily be absorbed into the other parameters, but it provides us with a convenient dial to select the fastest stable simulation parameters.

In this transformed reference frame, the control parameter front that induces phase-separation is stationary, and the material is advected across the front.  We choose to align the advection velocity with the $x$-axis: $\vec{u}=(u_x,0)$.  The control parameter front is then implemented by setting the parameters for $x<x_f$ to their separating values, and setting the parameters for $x\ge x_f$ to their mixing values.  For example, the model parameter $a(\vec{r},t)$ from \eqnref{eqn:parameter-a} implies the simulation parameter:
\begin{equation}
	a(\vec{r}) = \begin{cases}
a_S\,, & x < x_f \\
a_M\,, & x \ge x_f
\end{cases}\,.
\end{equation}

Derivation of the equation of motion from these moments can be accomplished by a Taylor expansion of the lattice Boltzmann equation (\ref{eqn:LBE}) with repeated substitution of the unknown functions $f_i$ with first order approximations in terms of the known equilibrium $f_i^0$.  This is shown in more detail in our previous paper for the one-dimensional system\cite{foard-2009}, and the analysis in higher dimensions remains essentially the same.  In terms of the simulation parameters, the equation of motion becomes:
\begin{equation}
\frac{1}{s}\partial_t\phi - \nabla \cdot \left( \vec{u} \phi \right) + O\left( \partial^3 \right) = \nabla \cdot \left( \tau - 1/2 \right) \nabla \mu \,.
\end{equation}
So we have to identify the mobility $m$ with $m=\tau-1/2$.  It turns out that this method does not show numerical artifacts for rapidly changing mobilities.  This suitibility for simulating abrupt changes in $\tau$ was previously used to simulate abrupt changes in dielectric properties of a medium\cite{wagner-may-2007}.

To fulfill Galilean invariance and isotropy requirements on a square lattice, implied by our choice of moment distributions Eqs.~(\ref{eqn:moments}) for an arbitrary advection velocity $\vec{u}(\vec{r},t)$--as would be the case were this simulation coupled to a hydrodynamic flow--would require the use of nine (or seven on a hexagonal lattice) velocity vectors in two dimensions.  However, because $\vec{u}=u_x$ is fixed and aligned to the $x$-axis we require only five velocities in two dimensions:
\begin{equation}
\vec{v}_i = \left\{ \cvec{0\\0},\cvec{-1\\0},\cvec{1\\0},\cvec{0\\-1},\cvec{0\\1} \right\}
\end{equation}
a so-called D2Q5 LB implementation.  For this velocity set we use the equilibrium distributions:
\begin{eqnarray}
\nonumber f_0^0 &=& \left( 1 - u_x^2 s^2 \right) \phi - 2 \mu s \,, \\
\nonumber f_1^0 &=& \frac{1}{2} \left( u_x^2 s^2 - u_x s \right) \phi + \mu s /2 \,, \\
\nonumber f_2^0 &=& \frac{1}{2} \left( u_x^2 s^2 + u_x s \right) \phi + \mu s /2 \,, \\
f_3^0 &=& f_4^0 = \mu s /2  \,.
\end{eqnarray}
In lattice Boltzmann units the material velocity is $- u_x s$, and can be made arbitrarily small by adjusting $s$.

The equilibrium distributions contain the chemical potential $\mu$ given by \eqnref{eqn:chem}.  The chemical potential contains a Laplacian which is evaluated on the discrete lattice as
\begin{equation}
	\nabla^2\phi(x,y) = \!\!\! \sum_{i,j=-1}^{1} \!\!\! w_{ij} \phi(x+i,y+j) \,,
\end{equation}
where the weights $w_{ij}$ are the elements of the stencil matrix
\begin{equation}
\vec{w} = \frac{1}{4}
\begin{pmatrix}
1 &   2 & 1 \\
2 & -12 & 2 \\
1 &   2 & 1
\end{pmatrix}\,,
\end{equation}
where the row and columns are numbered $\{-1,0,1\}$.  This choice of discrete Laplacian with $(x\pm1,y\pm1)$ terms is less susceptible to certain instabilities than those which have only $(x\pm1,y)$, and $(x,y\pm1)$ terms, allowing us to use an almost twice as large an effective time step\cite{pooley-2003} at very small computational cost.

We still need to define the boundary conditions. The $y$-dimension boundary conditions are periodic.  The inflow boundary condition at $x=l_x$ is straightforward with homogeneous material advancing with a constant prescribed flux:
\begin{equation}\label{boundary}
f_1(x=l_x-1,y,t+1) = f_2(x=l_x-1,y,t) + u_x s \phi_{in} \,.
\end{equation}
To calculate the Laplacian on this boundary we simply set $\phi(l_x,y)=\phi_{in}$.

The outflow boundary condition is somewhat more complicated, since we now have phase-separated material that is advected out and we need information about $\phi$ and $\mu$ from lattice sites that are not part of the simulation space.  We define our outflow boundary condition with the understanding that it should be neutral wetting to all concentration values, should not introduce gradients on the chemical potential, and any effect on the morphology should be short ranged compared to the simulation size.  This is accomplished by simply bouncing back the outflow density distribution after subtracting the material advected out of the simulation:
\begin{equation}
f_2(x=0,y,t+1) = f_1(x=0,y,t) - u_x s \phi(x=0,y,t) \,.
\end{equation}
For the purpose of calculating the Laplacian at the outflow boundary we set off-lattice concentrations to be the same as the boundary $\phi(-1,y)=\phi(0,y)$.  The effect of the outflow boundary condition on the bulk of the simulation can be ascertained by observing a normally stationary morphology as it is advected out of the simulation.  One choice is a single large circular region of $\mathcal{B}$-type material suspended in a bulk which is otherwise entirely $\mathcal{A}$-type.  Tests such as these (not shown) verify that this boundary condition has a very small effect, and is acceptable for this model.  Further evidence can be found by comparing phase-separation structures in \figref{fig:slow-early} and \figref{fig:fast-middle} which are nearly identical, as discussed later.

This fully describes our method which allows us to simulate the dynamics of structure generation by phase separation fronts.  As a final note: while simulations will show material moving past a stationary front, in our discussion of these simulation results we will always refer to the front as moving and the bulk of the material as stationary.

\section{First Survey}\label{scn:survey1}
Using our lattice Boltzmann method we now investigate how the phase separation front will influence the formation of structure.  To do this we set up a medium size simulation of $x=512$ by $y=1024$ lattice points. We put the position of the front at $x_f=384$ (in lattice units).  The region from $x=0$ to $x=x_f$ undergoes homogeneous phase-separation.  We use the initial condition $\Phi(\vec{r})=\Phi_{in}+0.01\xi$, where $\xi$ is random noise uniformly distributed in the range $[-1,1]$, as is typically done for homogeneous phase-separation simulations.  The other simulation parameters are: $a_M=1$, $a_S=-1$, $b_M=0$, $b_S=1$, $c_M=-\phi_{in}$, $c_S=0$, $m_M=m_S=1/2$, and $\kappa_M=\kappa_S=2$, consistent with the choice of non-dimensional parameters made at the end of \scnref{scn:model}.  The time scaling parameter $s=0.026$ is chosen as a maximum numerically stable effective time step for these parameters\cite{wagner-2007}.

\begin{figure}
	\begin{center}
		\subfigure[$T=82$, $UT=10.5$]{\label{fig:slow-early}\includegraphics[clip=true, width=0.31\linewidth]{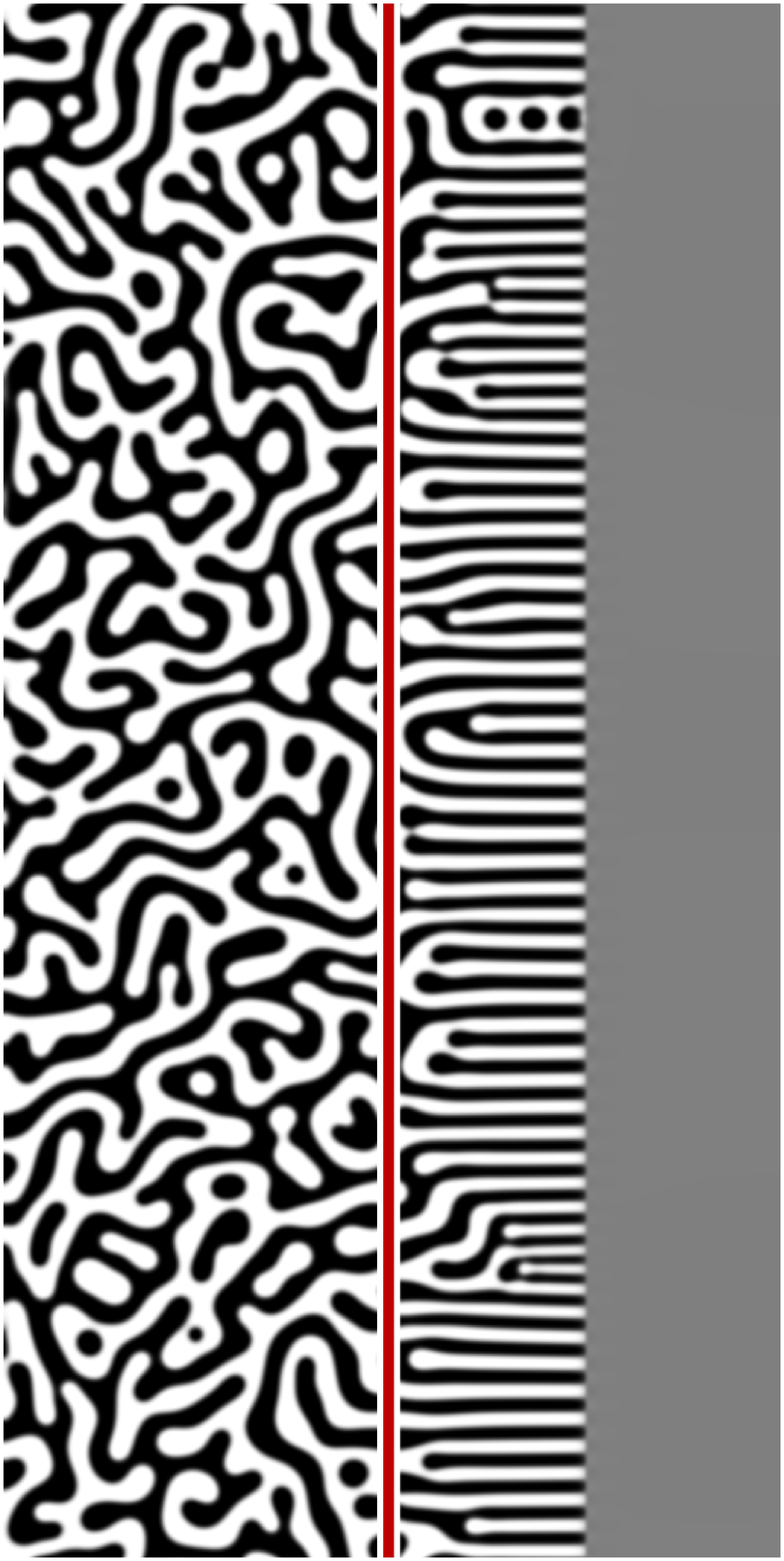}}
		\hfill
		\vline
		\hfill
		\subfigure[$T=164$, $UT=20.9$]{\label{fig:slow-middle}\includegraphics[clip=true, width=0.31\linewidth]{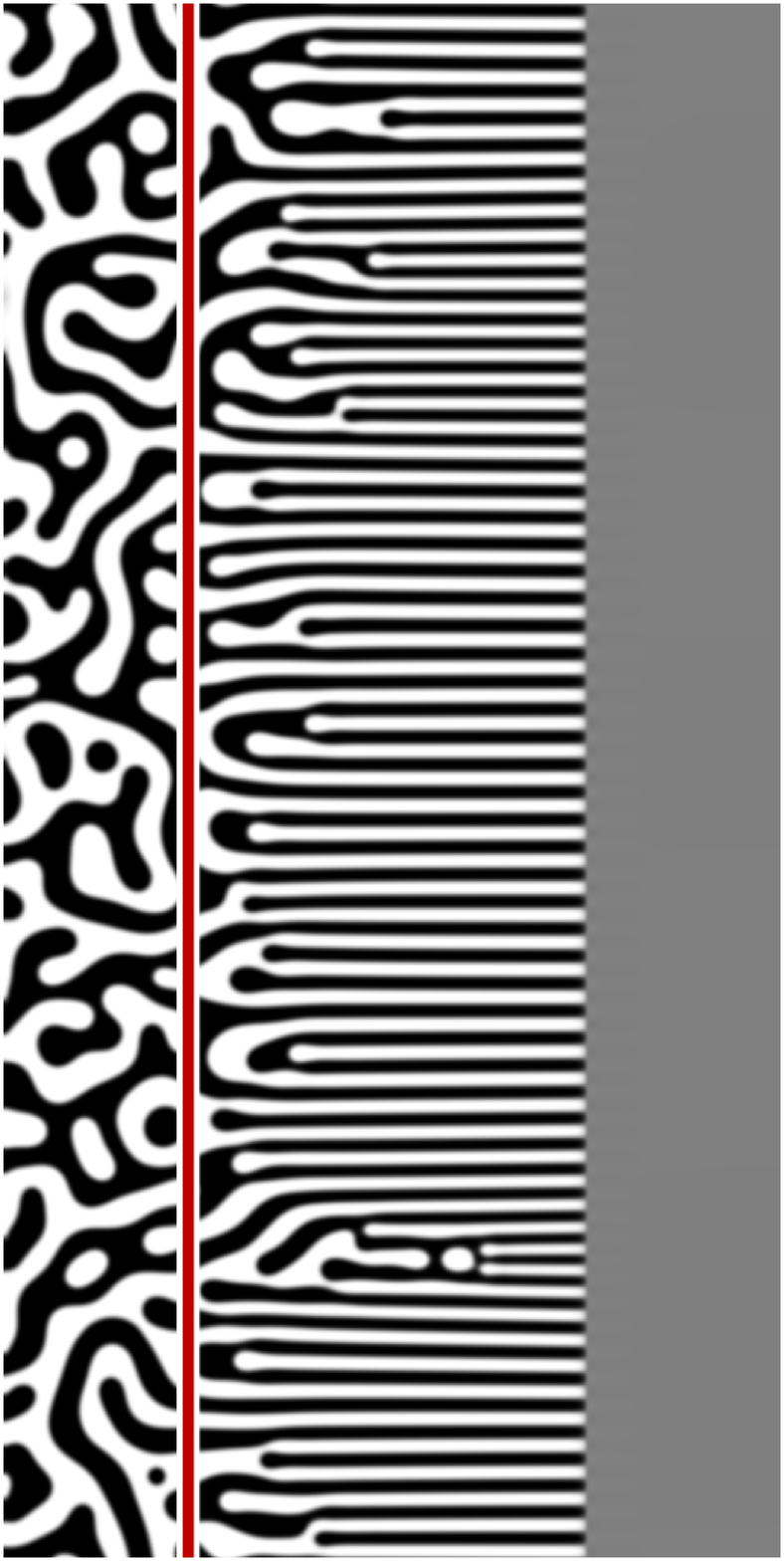}}
		\vline
		\hfill
		\subfigure[$T=966$, $UT=123.6$]{\label{fig:slow-late}\includegraphics[clip=true, width=0.31\linewidth]{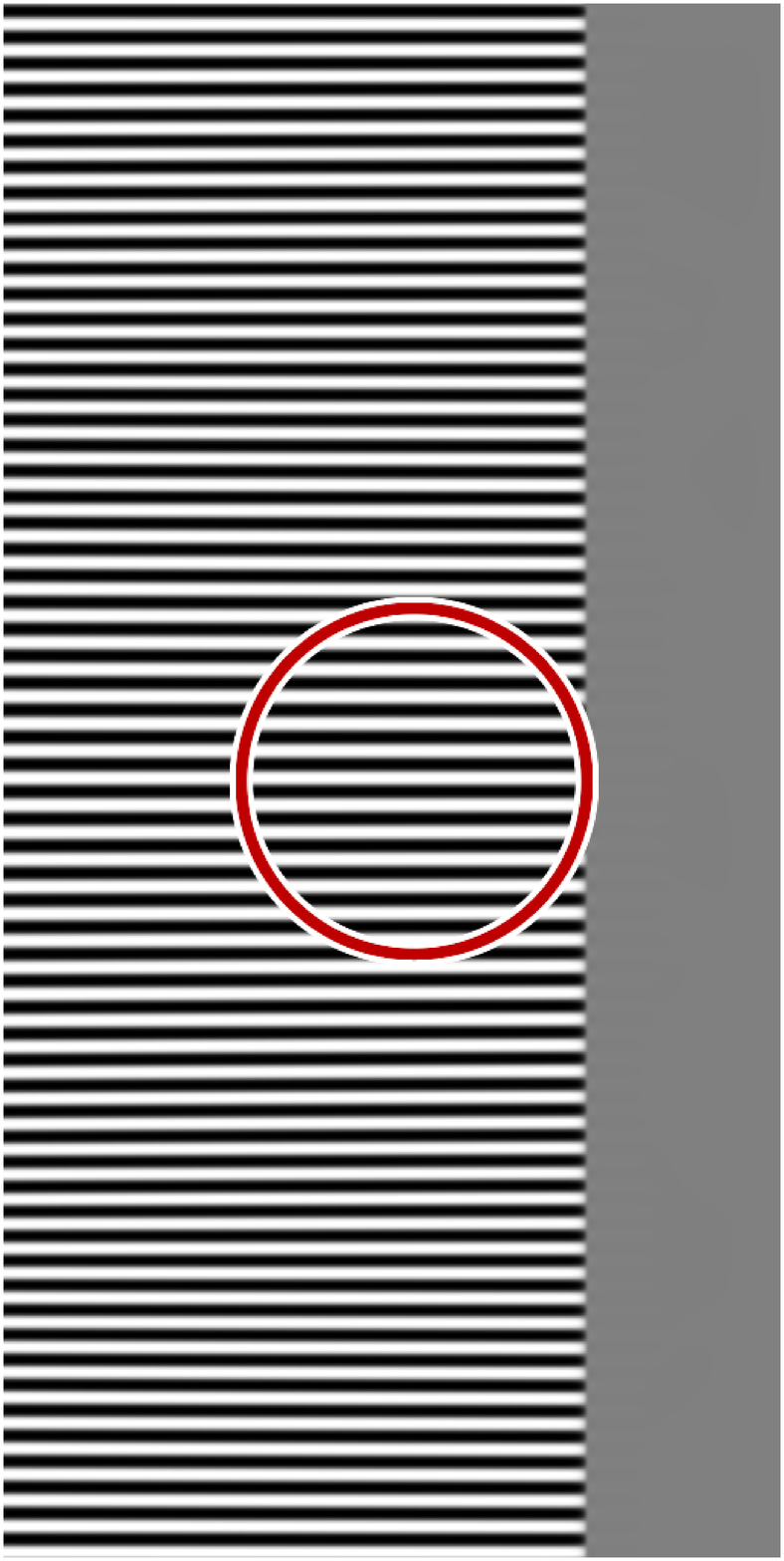}}
		\caption{\label{fig:slow}(Color online) Simulation showing stripe formation in the wake of a phase separation front.  For this simulation we choose a front speed $U=0.128$ and a symmetrically mixed $\Phi=0$ initial material concentration.  The starting location of the front is marked as a vertical stripe.  The rightmost image is the final structure observed after the front has moved far into the material.  A sample of the final morphology, noted by the circled region, will be compared to samples from other simulations performed with different parameter values.}
	\end{center}
\end{figure}

Such a simulation for $U=0.128$ is shown in \figref{fig:slow}.  The current position of the front at any time is easily visible as the transition between the black-and-white phase-separated region and the gray mixed region.  In \figref{fig:slow-early} the front has moved only a short distance of $x = 132 = 10.5 \,\lambda_{sp}$ (in units of the spinodal wavelength).  The position where the front started is marked by a vertical stripe.  The area to the left of the initial location of the front has undergone normal spinodal decomposition generating a phase-separation morphology typical of homogeneous phase-separation.  As the front moves on, new material phase-separates, but we see immediately that the structure of these newly formed domains is quite different from the domains formed through homogeneous phase-separation.  In the region between the initial front position and the current front position we observe a different kind of morphology: the domains are oriented orthogonally to the front, and show a variety of widths. In \figref{fig:slow-middle} the front has advanced a distance $x = 263 = 20.9 \,\lambda_{sp}$. The region of homogeneous phase-separation has noticeably coarsened and the newly overtaken material continues to phase-separate into a striped morphology. The striped structure, however, is not homogeneous. Particularly where the stripes are thin, defects can be seen to traverse into the striped domains. In \figref{fig:slow-late} the front has traveled a distance $x = 1553 = 123.6 \,\lambda_{sp}$. Now all the defects have been advected out of the system and the stripes are taking on a uniform thickness.   This is a stationary solution that will persist indefinitely.  Note that there is no evidence of any further coarsening at the position of the front.

\begin{figure}
	\begin{center}
		\subfigure[$T=33$, $UT=8.4$]{\label{fig:fast-early}\includegraphics[clip=true, width=0.31\linewidth]{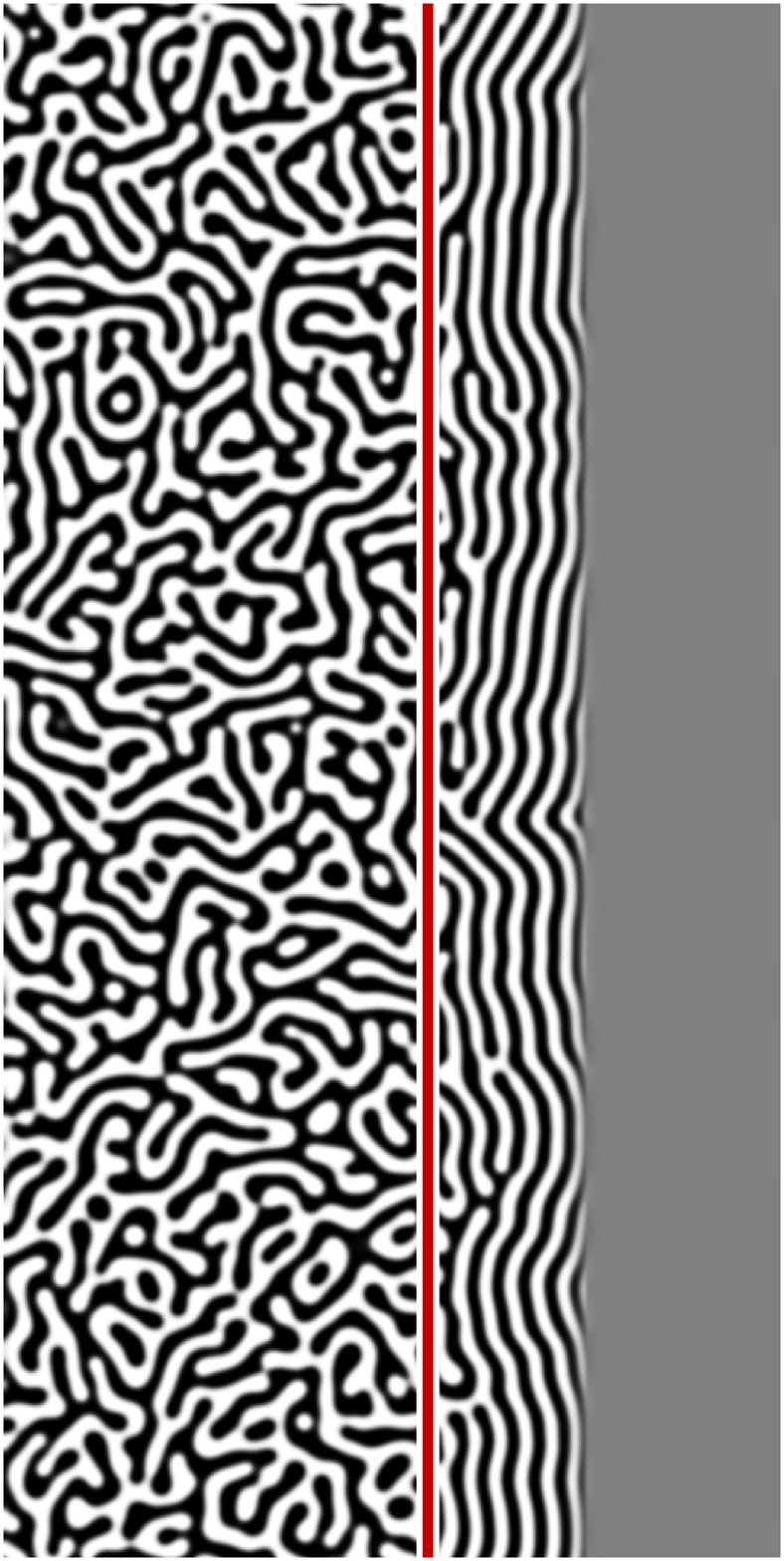}}
		\hfill
		\vline
		\hfill
		\subfigure[$T=82$, $UT=20.9$]{\label{fig:fast-middle}\includegraphics[clip=true, width=0.31\linewidth]{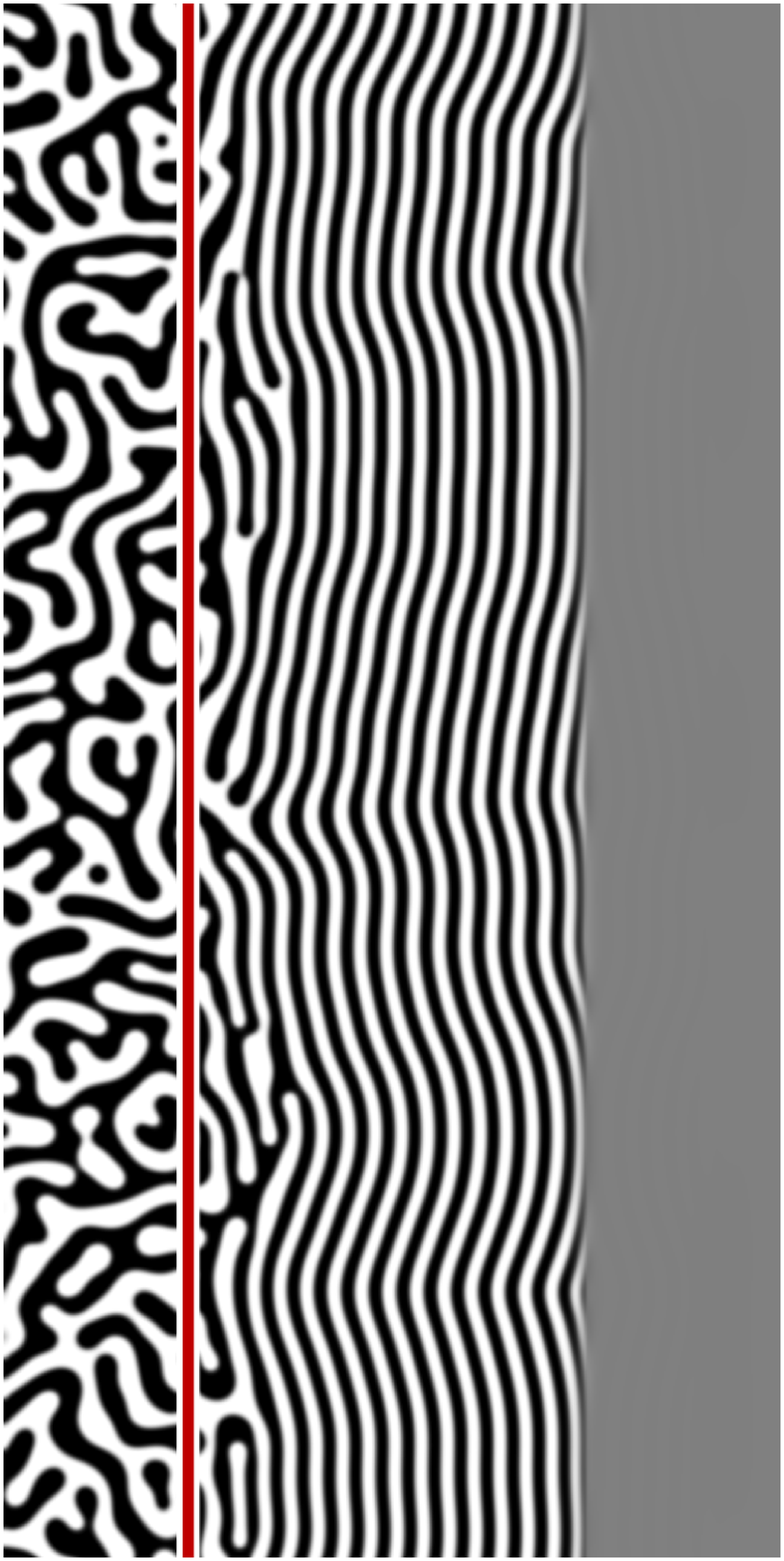}}
		\vline
		\hfill
		\subfigure[$T=491$, $UT=125.7$]{\label{fig:fast-late}\includegraphics[clip=true, width=0.31\linewidth]{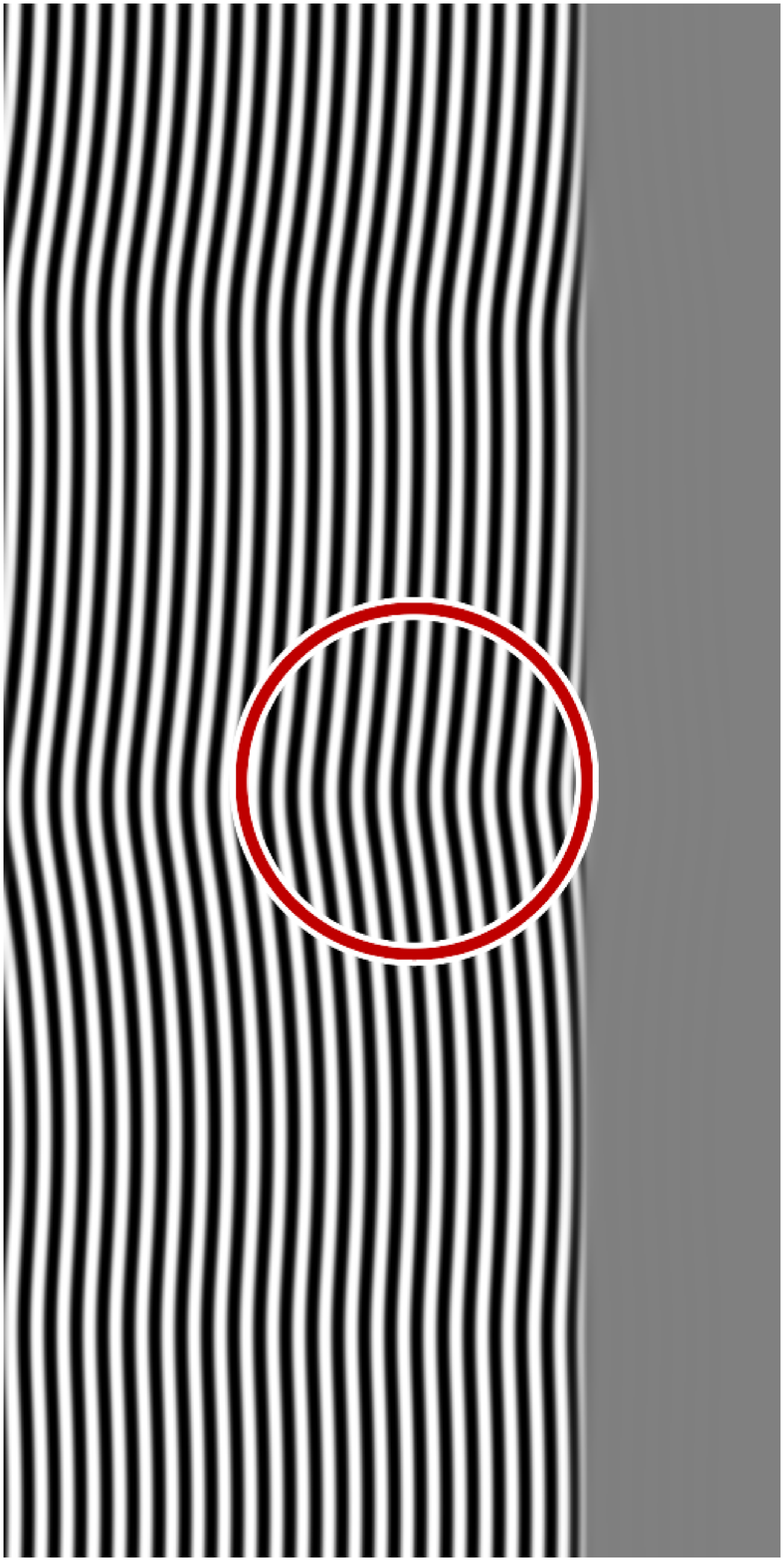}}
		\caption{\label{fig:fast}(Color online) Similar to \figref{fig:slow}, this sequence shows a simulation resulting in stripe formation by a phase separation front moving with speed $U=0.256$ into mixed material of initial concentration $\Phi=0$.  At this faster speed the stripes formed by the front are oriented parallel to the front.}
	\end{center}
\end{figure}

To examine the effect of front speed we now perform a simulation for $U = 0.256$.  The results of this simulation are shown in \figref{fig:fast}.  This time the morphology formation at the front is qualitatively different.  Again a regular striped morphology is formed, but it is now oriented parallel to the front in agreement with previous results\cite{furukawa-1990, hantz-2006}.  In \figref{fig:fast-early} the front has traveled a short distance of $x = 105 = 8.4\,\lambda_{sp}$. We observe typical homogeneous phase-separation morphology behind the original front location, however, where the front has traversed there are stripes of somewhat regular widths oriented roughly parallel to the front.  While the stripe widths are fairly uniform, there are still a large number of bends in these stripes.  In \figref{fig:fast-middle} we see that as time progresses new stripes form with fewer sharp bends, but the stripe widths do not appreciably change.

Note that the region of homogeneous phase-separation corresponds to the homogeneous phase-separation in \figref{fig:fast-early} at the same non-dimensional time. The initial noise on the order parameter was identical for both simulations and closer examination shows that the resulting phase-separation morphologies are nearly identical. This shows that there is little interaction between the regions of striped morphology and the homogeneous region. It also shows that neither the outflow boundary condition nor the advection speed significantly influence the simulations. This indicates that the coarsening of the region which separated under homogeneous conditions only slightly affects the striped morphology.

The width of stripes oriented parallel to the front can be understood by considering this as a quasi one-dimensional system. We analyzed this situation in an earlier paper\cite{foard-2009} and found that the wavelength of the parallel stripes follows from the front speed as
\begin{equation}
L(U) = \frac{8 \ln\left( 2 - \sqrt{2/3} \right) + 4 \sqrt{2/3} -4 }{\pi^2U} \,,
\end{equation}
for very slow fronts ($U < 0.001$) moving into material that has vanishingly low diffusive mobility ahead of the front ($M \rightarrow 0$).  For faster speeds this relation breaks down, and this theoretical prediction is inappropriate for the front speed of $U = 0.256$ in the example simulation shown in \figref{fig:fast}.  However, the observed wavelength of these quasi-one-dimensional stripes $L_{\parallel\text{2D}} = 1.39$ with $M=1$ compares favorably with the measured stripe wavelength $L_\text{1D} = 1.36$ from the $M=0$ simulation results in our earlier paper on phase separation fronts in one-dimensional systems\cite{foard-2009}.

These two qualitatively different morphologies were first described by Furukawa\cite{furukawa-1990}.  They were later rediscovered by Hantz and Bir\'o \cite{hantz-2006} and appear also to be related to structures formed from eutectic mixtures, although typically a phase-field formalism is used to describe these structures.

\begin{figure*}
  \begin{center}
  	\includegraphics[width=\textwidth,clip=true]{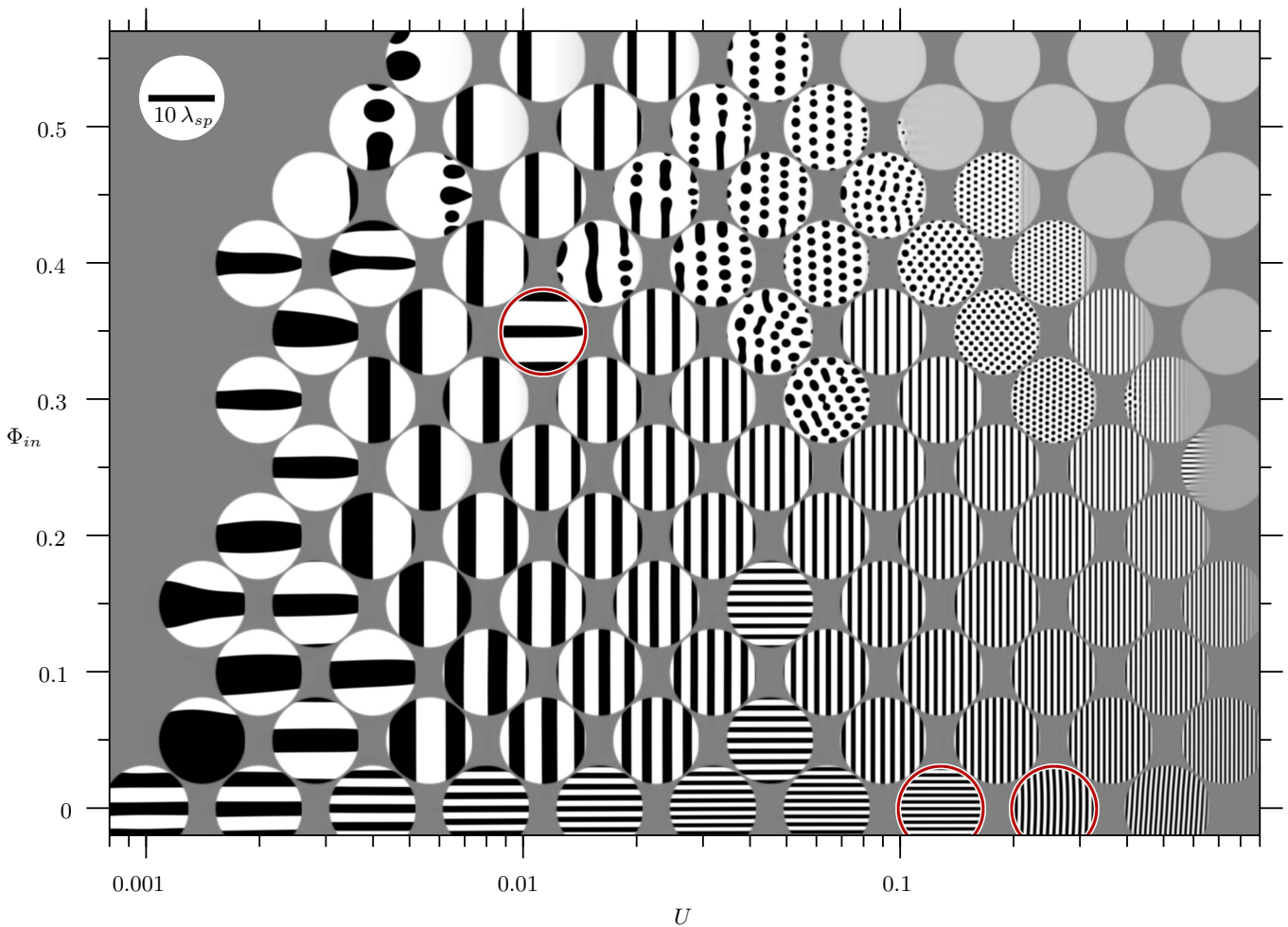}
		\caption{\label{fig:surveymap-random}(Color online) Morphology phase diagram from simulations started with random initial conditions.  Examples from this map are shown in \figref{fig:slow}, \ref{fig:fast}, and \ref{fig:depletion}.  The bar in the upper left corner has length 10 and height 1 in non-dimensional units.  The lack of an apparent pattern is due to the strong hysteresis of morphologies.  Further explanation is given in \figref{fig:depletion} and the text.}
  \end{center}
\end{figure*}

The fact that simply changing the velocity leads to a change in orientation of the domains raises the question as to where this transition happens, and surprisingly there appears to be no numerical value for the speed at which this transitions happens in the literature.  Also if we change the input composition, we will have to obtain stripes that have different width depending on the composition.  This more systematic investigation constitutes the main contribution of this paper.  Next we obtain a state-diagram from simulations such as the simulations as shown above.  We pick a sample of the morphologies--shown as a circle in \figref{fig:slow-late} and \figref{fig:fast-late}--and place these sample morphologies in a diagram at a position corresponding to $\Phi_{in}$ and $U$. We then performed simulations for a set of different values of $\Phi_{in}$ and $U$, but with all other parameters kept constant.  We ran each of these simulations until the front had moved approximately $1500$ lattice sites, or about $4$ times the distance from outflow boundary at $x=0$ to the front position at $x_f=384$.  We again use a circular section near the front, as indicated in \figref{fig:slow-late} and \figref{fig:fast-late}, and put them in a $\Phi_{in}$/$U$ graph at the appropriate position. This survey of resulting mophologies is shown in \figref{fig:surveymap-random}.

The result of the survey is initially surprising: while there is a clear transition from parallel to orthogonal stripes for $\Phi_{in}=0$ at around $U\approx 0.2$, orthogonal domains appear to be an anomaly only observed for exactly symmetric domains. However, there appears to be another transition between orientations of asymmetric domains at around a hundredth of this speed at $U\approx 0.003$.  The state diagram shows one additional boundary between regions of stripes and droplets for more off-critical mixtures.
These droplet structures appear to be preferred for larger speeds. For even larger speeds and more off-critical mixtures we see only mixed material, visible as gray disks. This means that the speed of a phase-separation front moving with constant speed into the mixture is smaller than the imposed speed of our front, and phase-separation is unable to keep up with our front.  The transition to a free front, to the accuracy of this survey, is unaffected by the initial conditions. For the parallel stripes the free front speed is given by\cite{vansaarloos-2003-386}
\begin{equation}
U_\text{free}(\Phi_{in})= \frac{1}{27\pi}\sqrt{2\left(34+14\sqrt{7}\right)\left(3-9\Phi_{in}^2\right)^3} \,.
\end{equation}

\begin{figure}
	\begin{center}
		\subfigure[$T=164$, $UT=1.8$]{\label{fig:depletion-early}\includegraphics[clip=true, width=0.31\linewidth]{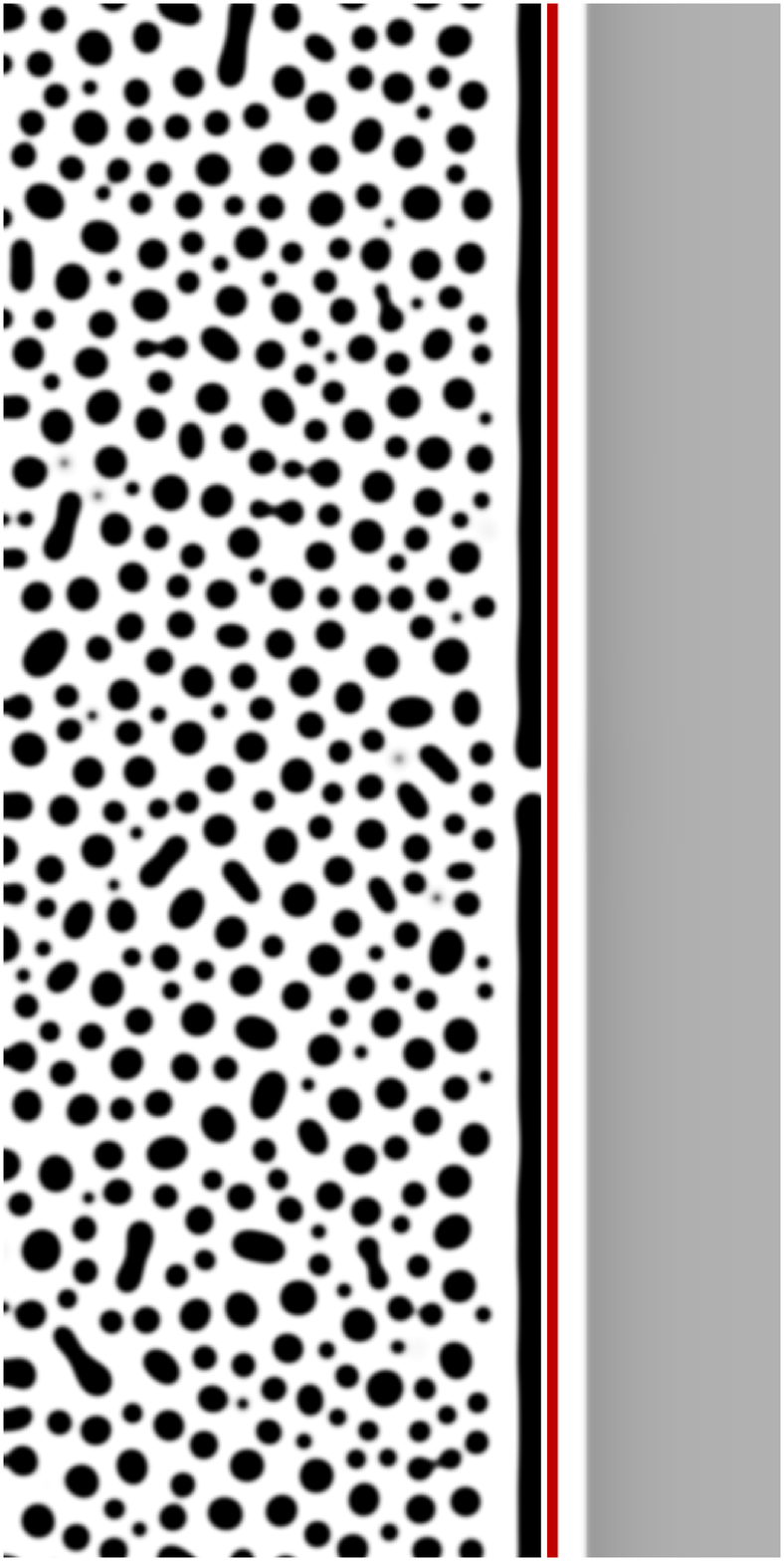}}
		\hfill
		\vline
		\hfill
		\subfigure[$T=2619$, $UT=29.6$]{\label{fig:depletion-mid}\includegraphics[clip=true, width=0.31\linewidth]{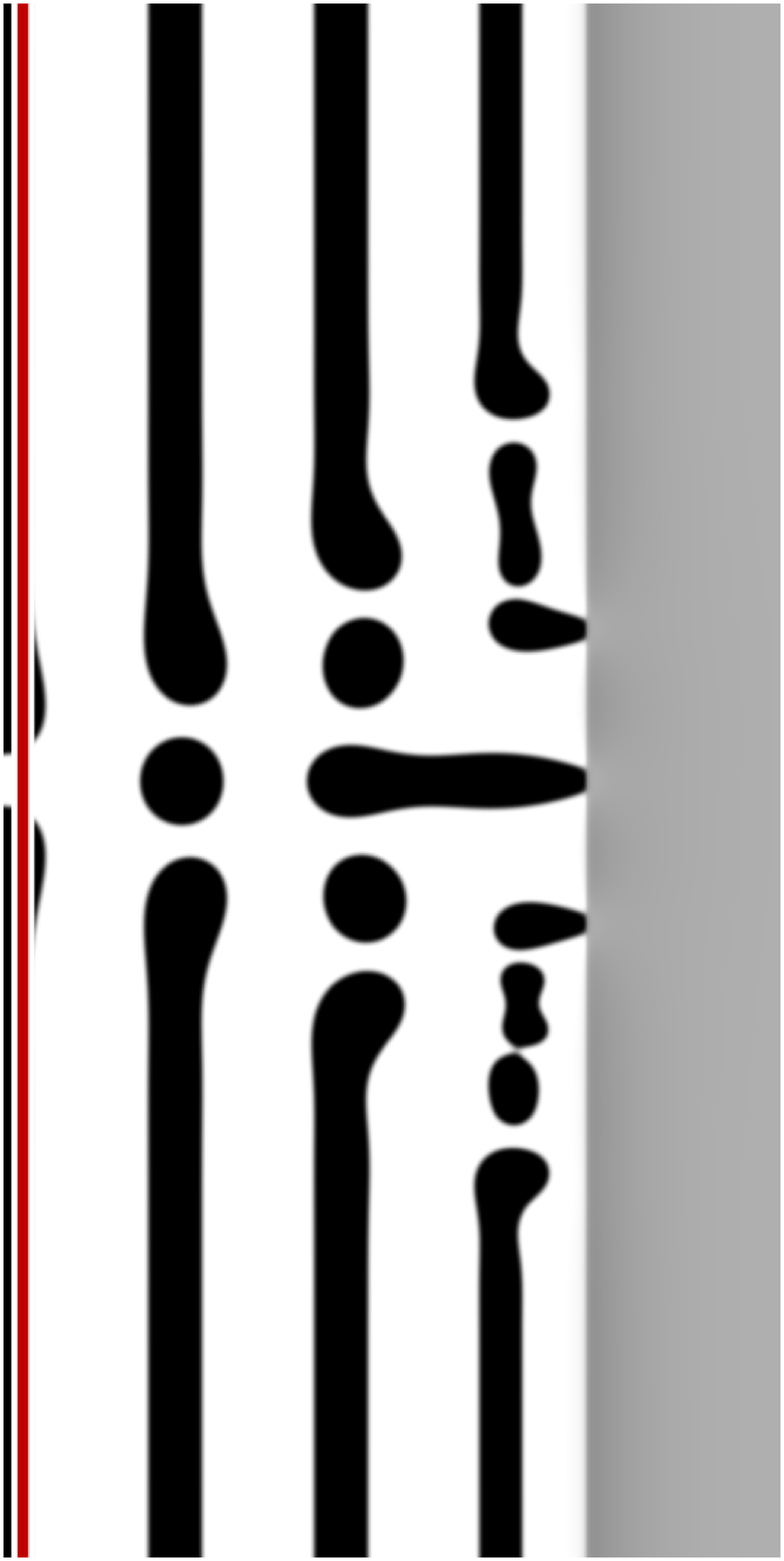}}
		\hfill
		\vline
		\hfill
		\subfigure[$T=5565$, $UT=62.9$]{\label{fig:depletion-end}\includegraphics[clip=true, width=0.31\linewidth]{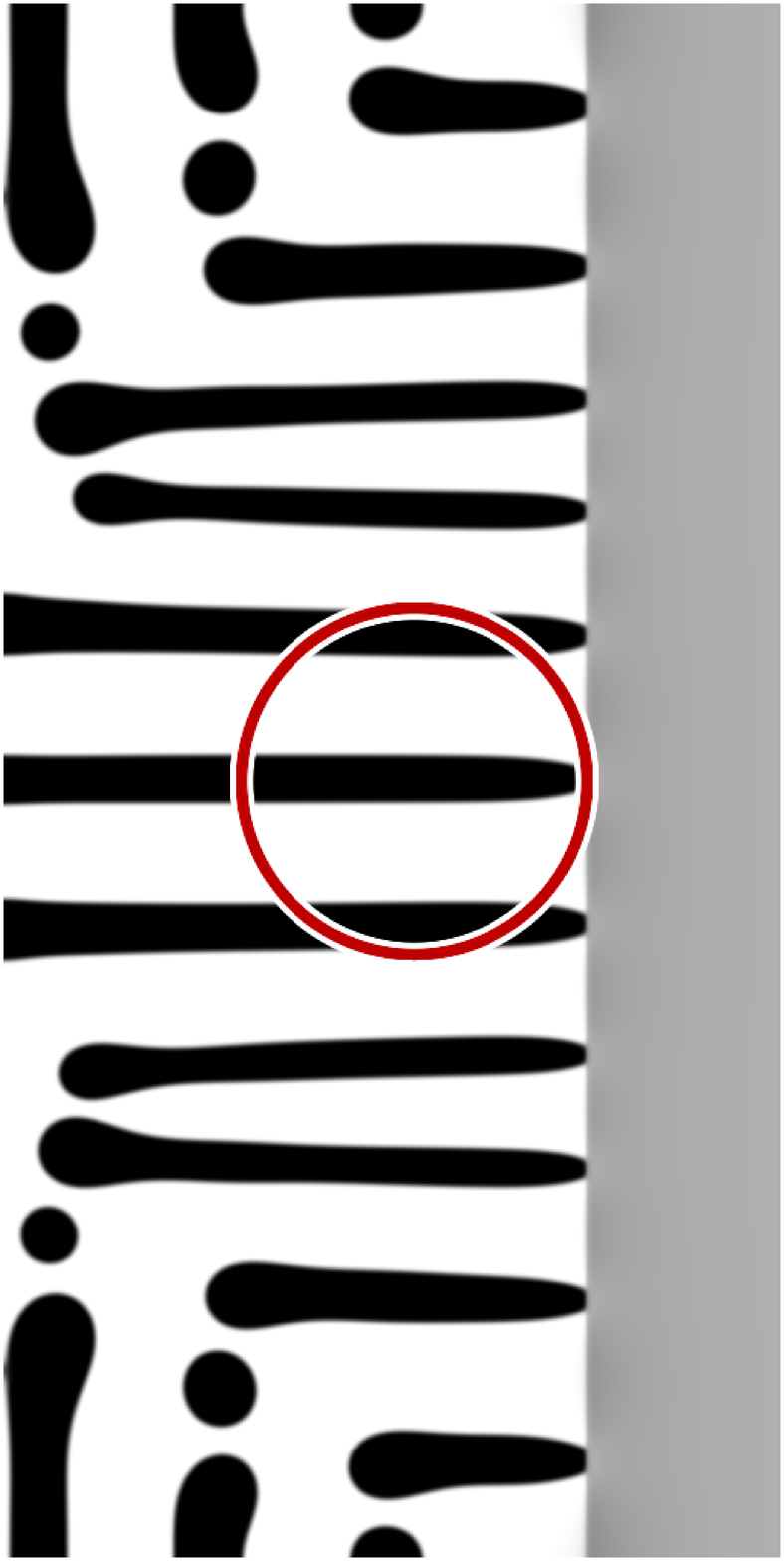}}
		\caption{\label{fig:depletion}(Color online) Simulation showing strong depletion effects.  The moving front suddenly appears in a mixed material with volume fraction $\Phi_{in}=0.35$ containing small fluctuations.  Shown in (a), the region far behind the front undergoes spinodal decomposition and droplet growth as in a homogeneous quench.  Near the front the formation of a depletion zone induces domains oriented parallel to the front.  By chance a defect in this domain facilitates a transition (b) to the favored orthogonal stripe morphology shown in (c).  The circle in (c) shows the region sampled for use in \figref{fig:surveymap-random}.  The front speed is $U=0.0113$ for this example.}
	\end{center}
\end{figure}

Closer examination of this state-diagram reveals more unexpected results. There are three examples of orthogonal stripes formed in a sea of parallel stripes at $(U,\Phi_{in})$ values of (0.045, 0.05), (0.045, 0.15), and (0.01, 0.35). These structures appear to break the prevalent trend of their neighbors and it is worth while to consider these simulations in some more detail. We will focus here on the simulation for ($U = 0.01$, $\Phi_{in}=0.35$). Three snapshots of this simulation are shown in \figref{fig:depletion}. The homogeneous spinodal morphology are droplets. However at the front a depletion zone favoring white material is formed.

The reason for this depletion layer is as follows: we chose to set the $C$-parameter in the mixing region such that $\mu(\Phi_{in})=0$ in the mixing region. After phase-separation we have $\mu(\Phi=\pm 1) = 0$ in the separated region and there will be no long range chemical potential gradient leading to extended diffusion of material into or out of the mixing region. Before the initial phase-separation in the separating region the order parameter is nearly uniformly $\Phi_{in}$ and $\mu(\Phi_{in})\neq 0$. Therefore we get some diffusion into and out of the mixing region leading, in this example, to a depletion of black material near the front. Once the phase-separation is complete, there is no longer a difference in the chemical potenial far away from the front.  

After the initial phase-separation with the creation of the depletion zone we then observe the nucleation of a first parallel black domain, as can be seen in \figref{fig:depletion-early}. So far this scenario is generic, but this simulation is special in that the formation of the first parallel stripe is not perfect, but a single defect was created. This defect now has a profound effect on the further evolution of the morphology formation. The next black domain that is formed has two defects with an interspersed drop. The following generation of parallel stripes has three drops, but the middle drop now maintains contact with the front, forming the first orthogonal stripe. The defect invades the formation of parallel domains leading to a wedge of orthogonal stripes that replace the parallel stripes. After a sufficiently long time we are left with a purely orthogonal morphology.

This suggests that instead of our state-diagram as shown in \figref{fig:depletion} we should associate the orientation of the domains with a probability, since the selection is apparently probabilitstic, depending on the details of the homogeneous phase-separation, which in turn depends on the initial noise. However, since the appearance of a single defect can be sufficient to switch the orientation (as shown in \figref{fig:depletion}) we expect the probability of finding a certain morphology to also depend on the system size, since it is more likely to develop such a defect in a larger system. In the limit of a macroscopic system, we would expect that the probability of finding a defect would increase significantly, so that it becomes more interesting to examine which morphology is the preferred morphology.

\section{Second Survey}\label{scn:survey2}
\begin{figure}
	\begin{center}
		\subfigure[$T=0$, $UT=0.0$]{\label{fig:stripe-1}\includegraphics[clip=true, width=0.47\linewidth]{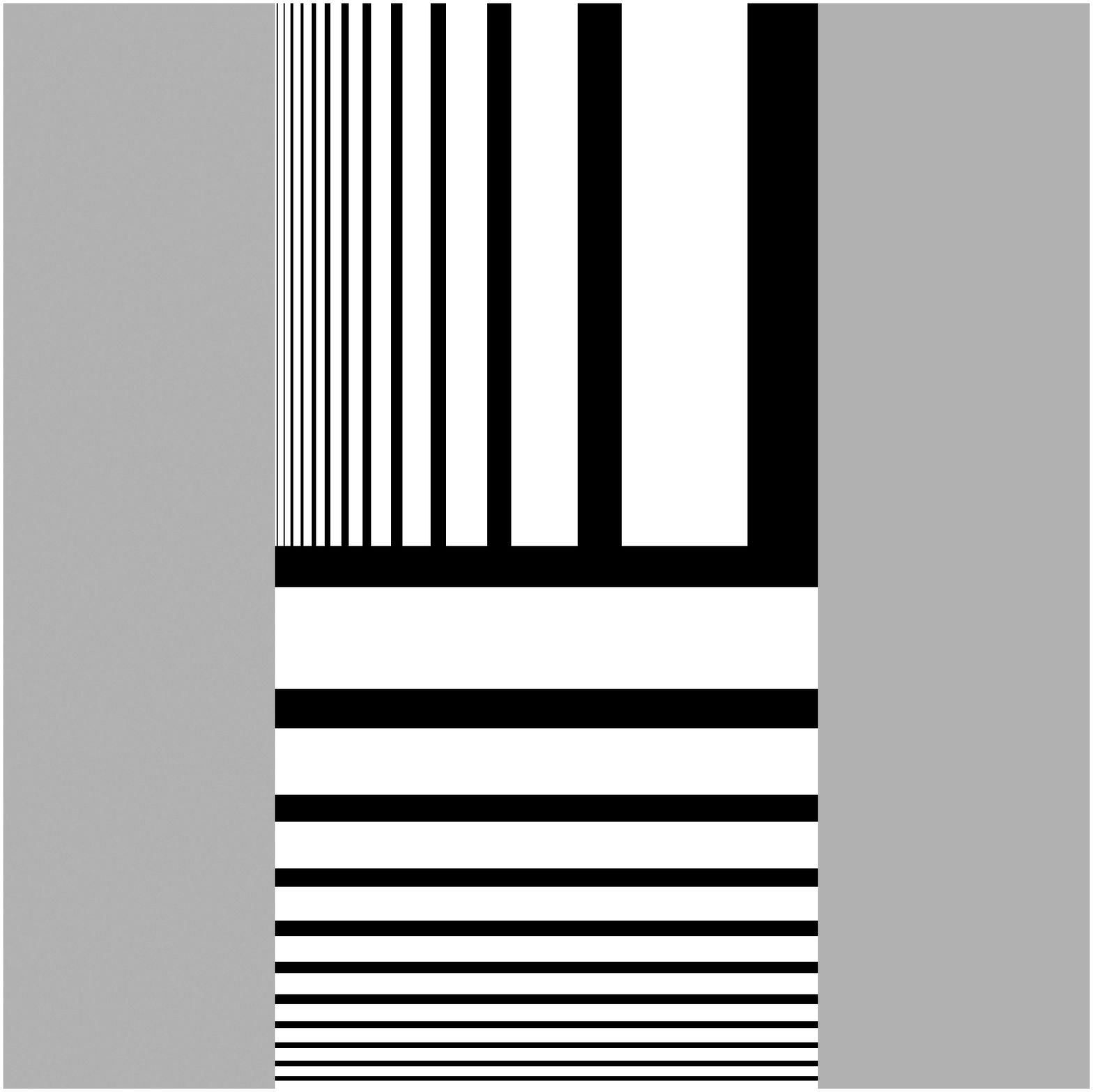}}
		\hfill
		\vline
		\hfill
		\subfigure[$T=573$, $UT=12.9$]{\label{fig:stripe-2}\includegraphics[clip=true, width=0.47\linewidth]{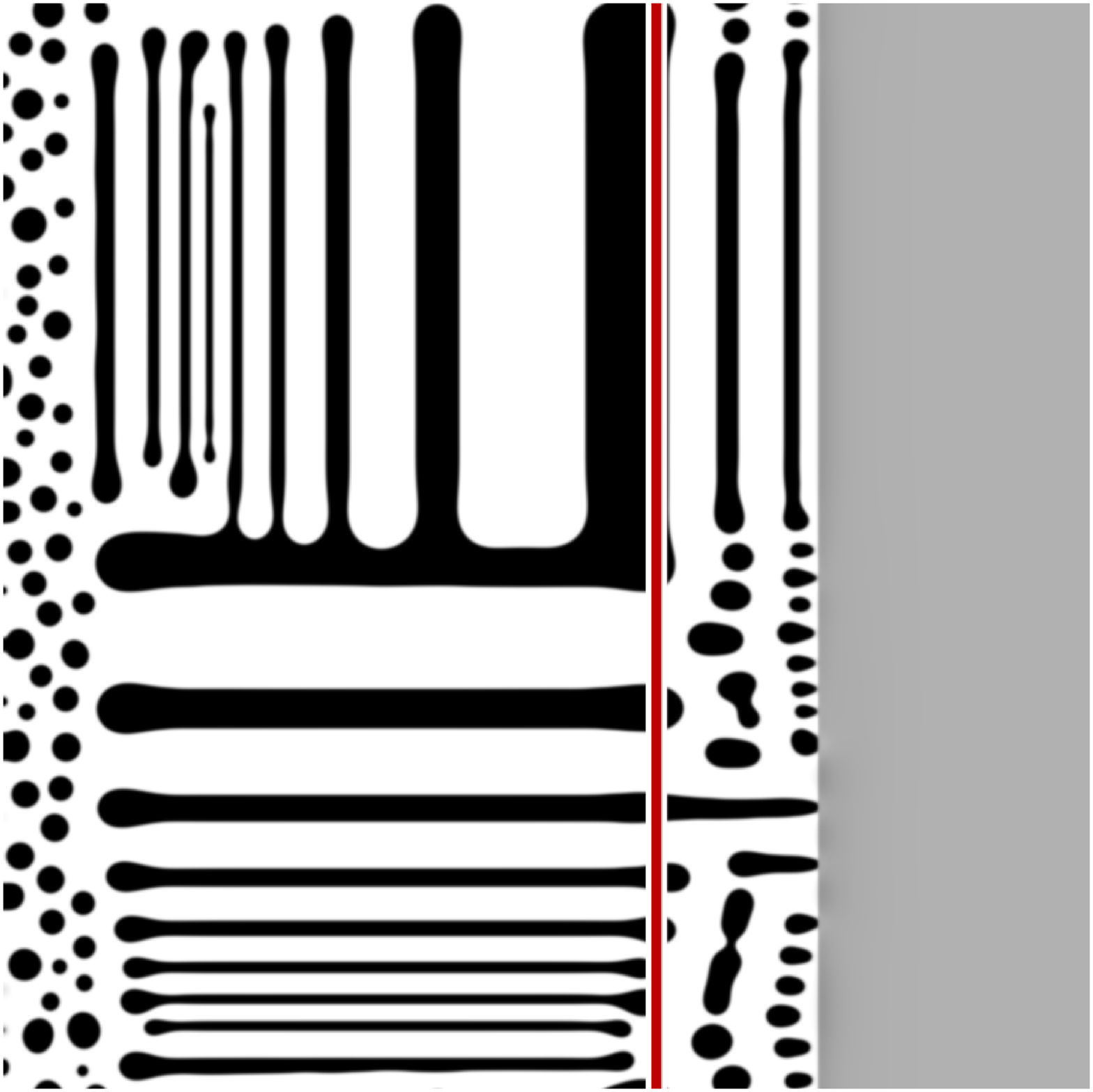}}
		\subfigure[$T=2292$, $UT=51.9$]{\label{fig:stripe-3}\includegraphics[clip=true, width=0.47\linewidth]{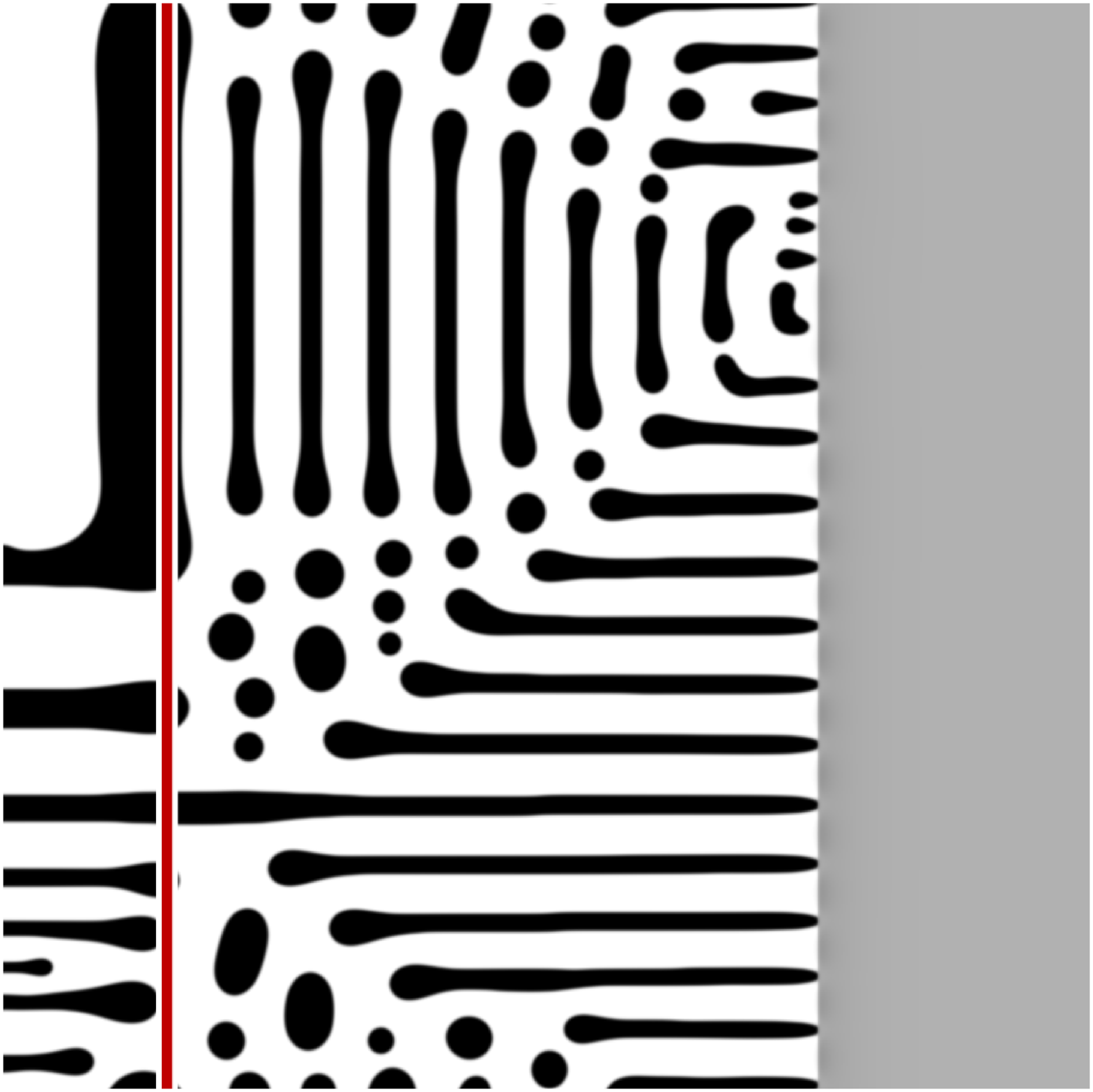}}
		\hfill
		\vline
		\hfill
		\subfigure[$T=9739$, $UT=220.4$]{\label{fig:stripe-4}\includegraphics[clip=true, width=0.47\linewidth]{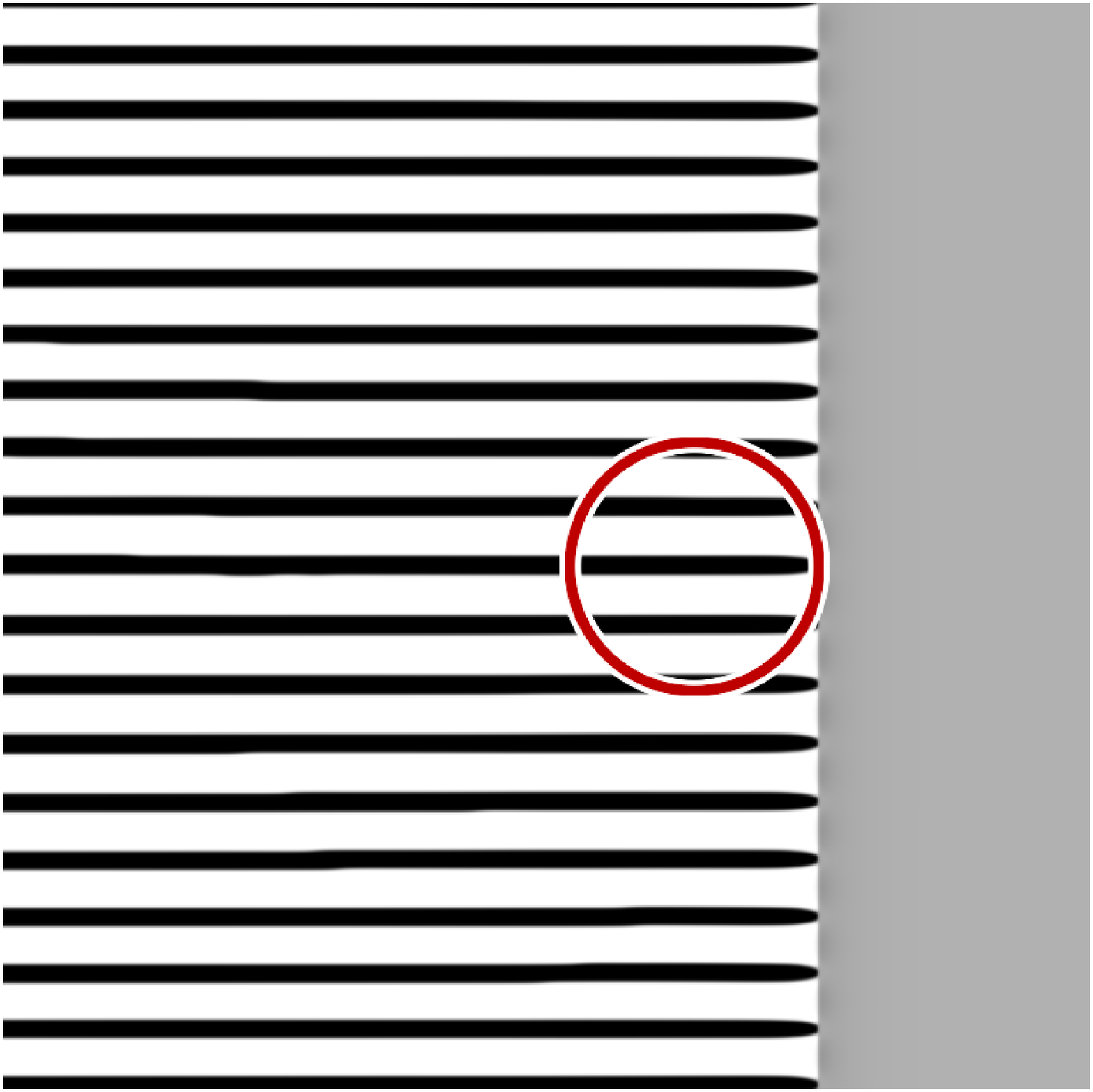}}
		\caption{\label{fig:stripe}(Color online) Example simulation showing selection of favored morphology for an imposed front speed $U=0.0226$ and initial concentration $\Phi_{in}=0.35$.  These parameters have a favored morphology of stationary stripes, but is close in parameter space to the droplet morphology region.  The initial phase separation configuration (a) allows for the selection of the favored morphology without the strong depletion effects observed in \figref{fig:depletion-early} which sometimes occur after spinodal decomposition near a front.  The stability of the periodic stripes is evident in (c).  The final configuration in (d) shows the stable stationary stripe morphology, and a circle which outlines the region sampled for use in \figref{fig:surveymap-selective}.}
	\end{center}
\end{figure}

To find out what is the preferred morphology we choose an initial condition which contains stripes of both parallel and orthogonal orientation and is consistent with the overall composition $\Phi_{in}$.  Parallel stripes are generated through a nucleation process and such a stripe selects its preferred length. For the formation of orthogonal strips no nucleation occurs and it is more difficult for such stripes to select a preferred length scale.  Here we design our initial condition with a range of stripe widths to allow for easy selection of the preferred width.  We show this initial condition in \figref{fig:stripe-1}.

We construct the initial conditions by selecting a square region of the simulation with side length $l$ that is half the simulation height.  Assuming this region has an origin in the lower left, the concentration in the region is given by:
\begin{equation}
	\Phi(x,y)=\sign \left[ \sin\left( \frac{2 \pi l y}{al - (a + b)y} \right) + \sin\left( \frac{\pi \Phi_{in}}{2}\right) \right] \;,
\end{equation}
where $a=\,\lambda_{sp}$ and $b=4\,\lambda_{sp}$ are, respectively, the smallest and largest stripe wavelengths initially generated.  The region described is the lower-center of \figref{fig:stripe-1}.  The upper-center region is constructed similarly, although with a $\pi/2$ rotation to favor formation of parallel oriented stripes.  To avoid the possibility of the outflow boundary condition interfering with stripe selection dynamics, we disconnect the initially phase separated regions from the boundary with a region of mixed material at the initial volume fraction containing small fluctuations similar to the region ahead of the front.

\begin{figure*}
  \begin{center}
  	\includegraphics[width=\textwidth,clip=true]{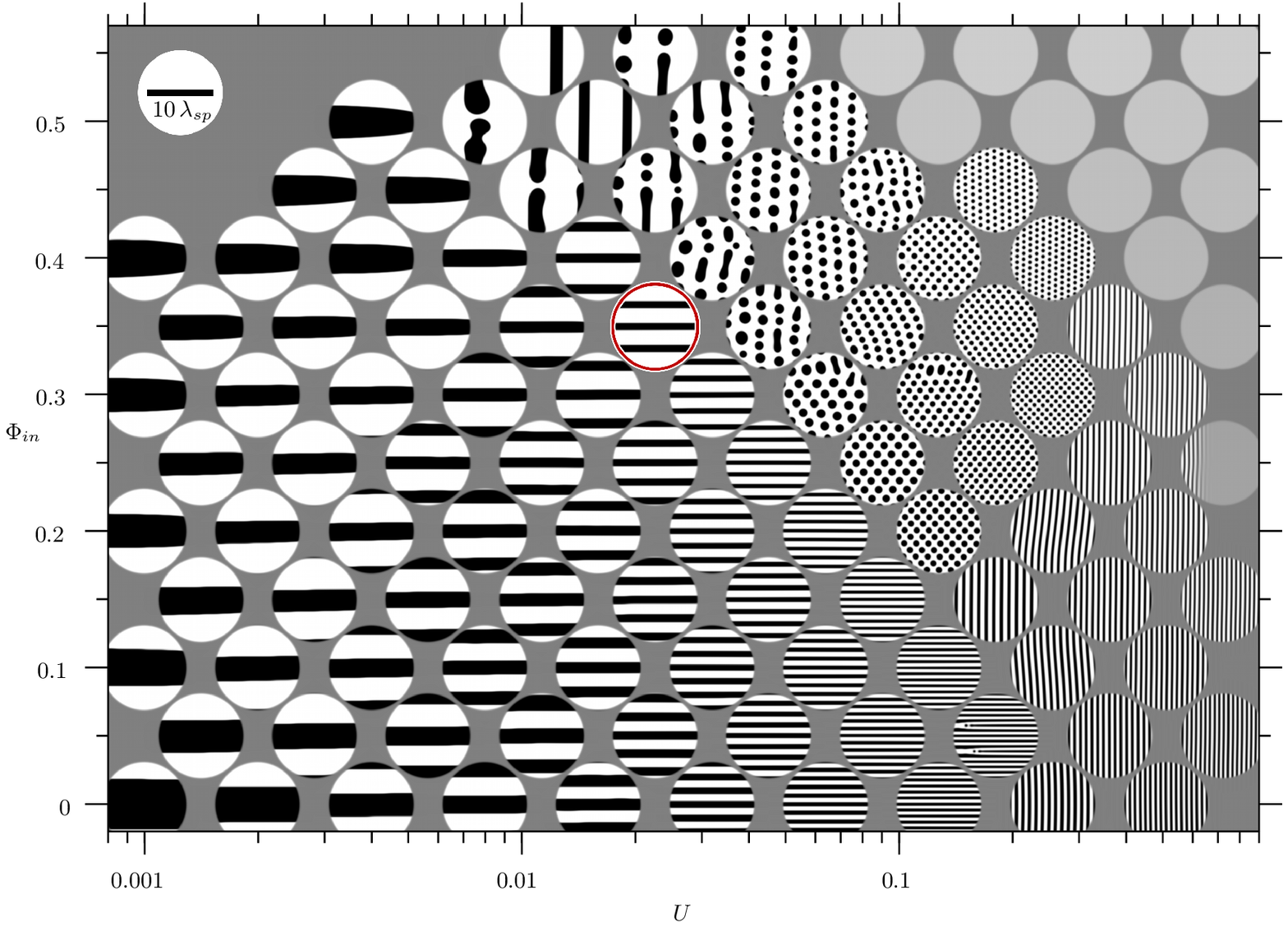}
		\caption{\label{fig:surveymap-selective}(Color online) Morphology phase diagram from simulations started with initial conditions containing a variety of morphologies.  The initial morphologies include long stripes oriented both parallel and orthogonal to the front with stripe wavelengths that range from one to ten spinodal wavelengths, and a region of mixed material with small fluctuations near the outflow boundary.  An example of the initial condition and final morphology selection is shown in \figref{fig:stripe}.}
	\end{center}
\end{figure*}

In \figref{fig:stripe-2} we see that only one of these initial stripes is selected to form the first orthogonal stripe. For this example the orthogonal stripes are again the preferred morphology (\figref{fig:stripe-2}) and the orthogonal stripes eventually replace the parallel stripes (\figref{fig:stripe-3}). Depending on the volume fraction and the front speed either orthogonal stripes (as shown here), parallel stripes, or droplets may turn out to be the preferred morphologies.

With this new initial condition we may now calculate a new state diagram.  The simulations used in this survey use the same non-dimensional parameters as the first survey.  Here the simulation size has been changed to $x=768$ by $y=1024$; widened to accommodate the new initial conditions.  To compensate for the increased computational cost of a larger simulation we chose to use a smaller value for the interfacial free energy cost $\kappa=1$, which increases the effective simulation speed by allowing a larger effective time step $s=0.079$\cite{pooley-2003}.  This new state diagram is shown in \figref{fig:surveymap-selective}. As expected we find that for the preferred morphologies there is now a much larger range of parameter values for which we find orthogonal stripes. For symmetric mixtures the transition between orthogonal and parallel stripes appears unchanged up to the accuracy of this survey. As we change the $\Phi_{in}$ from symmetrical ($\Phi_{in}=0$) to asymmetrical ($\Phi_{in}\neq 0$) compositions we now observe a continuous transition in morphologies. This is because we start with a phase-separated morphology in the separating region, and we no longer form a depletion layer. Thus parallel or orthogonal stripes are not \emph{a priori} favored. We see that increasing the volume fraction still lowers the speed for which we see a transition between orthogonal and parallel stripes, albeit in a much less drastic fashion.

\begin{figure}
	\begin{center}
		\psfrag{U}{$U$}
		\psfrag{0.001}{$0.001$}
		\psfrag{0.01}{$0.01$}
		\psfrag{0.1}{$0.1$}
		\psfrag{1}{$1$}
		\psfrag{Fi}{$\Phi_{in}$}
		\psfrag{MF}{$-1$}
		\psfrag{MM1M3}{$-\sqrt{1/3}$}
		\psfrag{0}{$0$}
		\psfrag{M1M3}{$\sqrt{1/3}$}
		\psfrag{F}{$1$}
		\psfrag{orthogonal stripes}{orthogonal stripes}
		\psfrag{parallel stripes}{parallel stripes}
		\psfrag{drop}{drop}
		\psfrag{array}{array}
		\psfrag{drop}{drop}
		\psfrag{array}{array}
		\psfrag{free front (periodic)}{free front (periodic)}
		\psfrag{free front}{free front}
		\psfrag{free front}{free front}
		\psfrag{(single domain)}{(single domain)}
		\psfrag{(single domain)}{(single domain)}
		\includegraphics[clip=true, width=\linewidth]{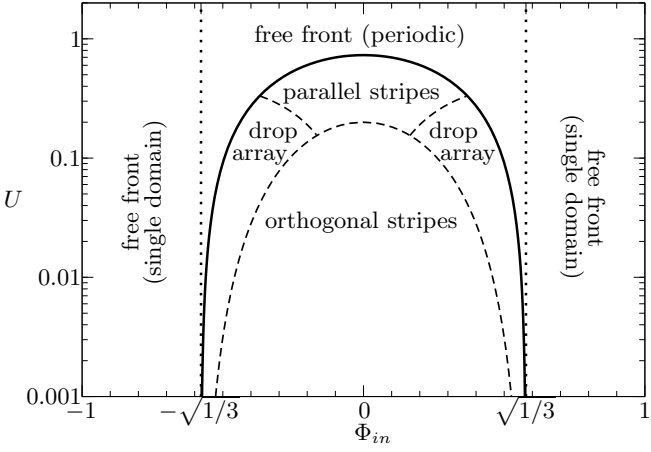}
		\caption{\label{fig:surveymap-morphology}Morphology state diagram of front induced phase separation generated structures in two dimensions. See text at the  end of \scnref{scn:survey2} for a detailed description of the regions and boundaries.}
	\end{center}
\end{figure}

We also find a larger parameter range for which we find droplet arrays, particularly for larger speeds and more symmetrical volume fractions.  We can now write a schematic state diagram for the preferred morphologies. This is shown in \figref{fig:surveymap-morphology}.
The region labeled ``free front (periodic)'' is where phase separation lags behind the control parameter front.  The ``free front (single domain)'' regions are where initially undifferentiated material will not spontaneously demix without fluctuations to induce nucleation.  Without fluctuations, instead of new domain formation, any existing domains of phase-separated material will slowly grow into the mixed material.  How structures which may form in these regions are effected by the passage of a control parameter front is beyond the scope of this paper.  The regions under the solid curve are based on observations of the preferred morphology evidenced in Figs.~\ref{fig:surveymap-random} and \ref{fig:surveymap-selective}.  Apart from the free front region we also show the regions where we observe parallel stripes, orthogonal stripes, and droplet arrays.  The former two we have covered in previous sections, but the later requires more discussion.  The droplet structure is observed to initially form near the front with little long-range order.  As the front progresses, the position of the newly forming droplets is influenced by the depletion of material caused by the formation of the previous drops.  The larger the drop formed, the more it depletes the surrounding region, and the further away the next droplets will form.  This mechanism causes reordering and elimination of small droplets in favor of larger droplets.  A nucleation condition imposes a maximum droplet size for a given front speed and mixed material concentration.  The result is a droplet structure which converges towards a highly ordered array of homogeneous size droplets.  Whether the droplet array has a preferred orientation, or if it is periodic in the reference frame of the front, are open questions.

\section{Analysis of Results for Critical Concentrations}\label{scn:analysis}
We have observed in the surveys that a phase separation front moving into material that is at the critical concentration $\Phi_{in}=0$ will form a striped morphology that is oriented either parallel or orthogonal to the front.  The parallel stripes are an essentially one-dimensional morphology, and have been discussed in our previous paper\cite{foard-2009}.  As mentioned in \scnref{scn:survey1}, the key property of parallel stripes is that for a given set of parameters there is a unique stripe wavelength, and the wavelength scales $L_\parallel \propto1/\sqrt{U}$.  By contrast, orthogonally oriented stripes may form with a wide range of wavelengths for a given parameter set.  In this section we analytically determine the stable wavelengths for orthogonal stripes as a function of the enslaved front speed.  We accomplish this in two stages.  We first determine the maximum wavelength of a stripe before a new stripe nucleates in its center, splitting it.  We next find that there is a metastable minimum stripe width, below which coarsening dynamics can result in stripes which disappear into the front.  The result is a region of stability for the orthogonal stripe morphology, outside of which we predict the favored morphology is parallel stripes or droplets.

\subsection{Maximum Orthogonal Stripe Size}
The mechanism responsible for limiting the maximum stripe wavelength is the nucleation of an opposite-type stripe at its center.  This can occur, even without fluctuations, due to the build-up of a nucleation kernel ahead of the enslaved front, similar to what happens in the one-dimensional case\cite{foard-2009}.

Formation of stable orthogonal stripes by a moving front results in a chemical potential profile that is stationary in the reference frame of the font.  We analytically determine this profile directly from the stationary solution ($\partial_T \Phi \rightarrow 0$) to the equation of motion:
\begin{equation}\label{eqn:poisson}
	0 = 
	\frac{1}{2\pi^2} \nabla_{\!\vec{R}}^2 \mu
	- \nabla_{\!\vec{R}} \cdot \left( \Phi \vec{U} \right) \;.
\end{equation}
The morphology under consideration is an array of highly ordered orthogonal stripes of wavelength $L_\bot$ with an $\mathcal{A}$-type stripe centered on the $X$-axis and the phase separation front on the $Y$-axis.  The front velocity $\vec{U}=(U,0)$ is entirely in the $X$-direction and constant, simplifying the gradient term to a derivative of the concentration $\Phi$ in the $X$-direction which is highly dominated by the concentration gradient at the front.  We approximated this term as an abrupt step which we expect to become exact in the limit of large stripes and slow fronts.  We focus on critical mixtures here, since additional complications occur for off-critical situations, as mentioned below.  For critical mixtures, the gradient term then simplifies to
\begin{equation}
	\label{eqn:sqrassume}
	\vec{\nabla}_{\!\vec{R}} \cdot \left( \vec{U} \Phi \right) = U \frac{d}{dX}\Phi \approx U \delta(X) \sqr(Y/L_\bot) \;,
\end{equation}
where $\sqr(x) \equiv \sign[\cos(2 \pi x)]$ is the square-wave function.  This is the only approximation made in our calculation.

An alternate form of the square-wave function is as an infinite series of delta function convolutions.  To make the series converge faster, each term of the series is centered on a crest and extends to include half of each adjacent trough:
\begin{equation}
	Sq(x) = \sum\limits_{n=-\infty}^{\infty} \left(
	2 \!\!\!\!\int\limits_{n-1/4}^{n+1/4}\!\!\!\! \delta(\widetilde{x}-x) \,d\widetilde{x} 
	- \!\!\!\!\int\limits_{n-1/2}^{n+1/2}\!\!\!\! \delta(\widetilde{x}-x) \,d\widetilde{x} \right) \;.
\end{equation}
Combined with the delta function property $\delta(x/\alpha)=\abs{\alpha}\delta(x)$ and the $n$-dimensional vector delta function identity $\delta_{n}(\vec{r})\equiv\delta(x_1)\delta(x_2)\cdots\delta(x_n)$, this alternate form allows \eqnref{eqn:poisson} to be rewritten
\begin{align}
	\nonumber
	\vec{\nabla}_{\!\vec{R}}^2 \mu
	= {} &
	2 \pi^2 U L_\bot \!\!\sum\limits_{n=-\infty}^{\infty}\!\! \left(
	2 \!\!\!\!\int\limits_{n-1/4}^{n+1/4}\!\!\!\! \delta_2(L_\bot\widetilde{Y}\vec{e}_Y - \vec{R}) \,d\widetilde{Y}
	\right. \\ & \left.
	- \!\!\!\!\int\limits_{n-1/2}^{n+1/2}\!\!\!\! \delta_2(L_\bot\widetilde{Y}\vec{e}_Y - \vec{R}) \,d\widetilde{Y} \right) \;,
\end{align}
where $\vec{e}_Y$ is the $Y$-axis unit vector.  The boundary condition for this morphology is a vanishing chemical potential far ahead and behind the front: $\mu(X\rightarrow\pm\infty)=0$. The Poisson equation $\nabla^2 f(\vec{r},\vec{r}_0) = \delta_2(\vec{r}_0-\vec{r})$ has the fundamental solution $f(\vec{r},\vec{r}_0) = \frac{1}{2\pi}\ln \abs{\vec{r}_0-\vec{r}}$.  This can be directly applied to the above equation to arrive at an infinite series integral solution for the chemical potential profile:
\begin{align}
	\nonumber
	\mu(L_\bot \vec{R})
	= {} &
	\pi U L_\bot \!\!\sum\limits_{n=-\infty}^{\infty}\!\! \left(
	2 \!\!\!\!\int\limits_{n-1/4}^{n+1/4}\!\!\!\! \ln \sqrt{X^2 + (\widetilde{Y}-Y)^2} \,d\widetilde{Y}
	\right. \\ & \left.
	- \!\!\!\!\int\limits_{n-1/2}^{n+1/2}\!\!\!\! \ln \sqrt{X^2 + (\widetilde{Y}-Y)^2} \,d\widetilde{Y} \right) \;,
\end{align}
It is easily seen that, by symmetry, this solution satisfies the chemical potential boundary condition.
The integrals are straightforward, however the result is somewhat lengthy:
\begin{widetext}
\begin{align}
\label{eqn:analytical}
	\nonumber
	\mu(L_\bot \vec{R}) = {} & \pi U L_\bot
	\!\!\sum\limits_{n=-\infty}^{\infty}\!\!
	\left\{
		(n - Y)
		\left[
			2 \arctanh{\frac{(n-Y)/2}{X^2+(n-Y)^2+1/16}}
			- \arctanh{\frac{n-Y}{X^2+(n-Y)^2+1/4}}
		\right]
	\right.
	\\ & \nonumber
	\left.
		+ X
		\left[
			  2 \arccot{\frac{X}{n+1/4-Y}}
			- 2 \arccot{\frac{X}{n-1/4-Y}}
			-   \arccot{\frac{X}{n+1/2-Y}}
			+   \arccot{\frac{X}{n-1/4-Y}}
		\right]
	\right.
	\\ &
	\left.
		+ \frac{1}{2} \ln
		\left(
			\frac{\left[ X^2 + (n+1/4-Y)^2 \right]\left[ X^2 + (n-1/4-Y)^2 \right]}
			{16 \left[ X^2 + (n+1/2-Y)^2 \right] \left[ X^2 + (n-1/2-Y)^2 \right]}
		\right)
	\right\}\;.
\end{align}
\end{widetext}

\begin{figure}
	\begin{center}
		\psfrag{XOL}{$X/L_\bot$}
		\psfrag{YOL}{$Y/L_\bot$}
		\psfrag{mOUL}{$\mu/U L_\bot$}
		\psfrag{M1O2}{$-1/2$}
		\psfrag{M1O4}{$-1/4$}
		\psfrag{0}{$0$}
		\psfrag{1O4}{$1/4$}
		\psfrag{M1O4}{$-1/4$}
		\psfrag{0}{$0$}
		\psfrag{1O4}{$1/4$}
		\psfrag{1O2}{$1/2$}
		\psfrag{3O4}{$3/4$}
		\psfrag{M2}{$-2$}
		\psfrag{M1}{$-1$}
		\psfrag{0}{$0$}
		\psfrag{1}{$1$}
		\psfrag{2}{$2$}
		\psfrag{Analytical Solution}{Analytical Solution}
		\psfrag{Simulation Result}{Simulation Result}
		\includegraphics[clip=true, width=\linewidth]{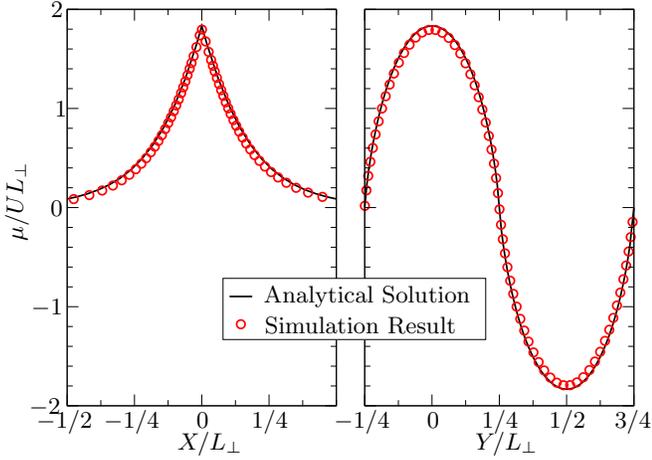}
		\caption{\label{fig:mu_profile}(Color online) Comparison of the analytical chemical potential profile of \eqnref{eqn:analytical} (line) with LBM simulation results (circles) for large stripes $L_\bot=70$ formed by a slow $U=0.00235$ front.  The left figure shows $\mu$ along the centerline of a stripe of $\mathcal{B}$-type material.  The right figure shows $\mu$ at the position of the front ahead of two adjacent stripes.  The profile in each figure intersect at $X/L_\bot=Y/L_\bot=0$.}
	\end{center}
\end{figure}
We verify the analytical chemical potential in \eqnref{eqn:analytical} by comparison to the chemical potential measured from a LBM simulation of a stripe pair with a large stripe wavelength $L_\bot=70$ and slow front speed $U=0.00235$.  The result of this comparison is shown in \figref{fig:mu_profile}, and demonstrates excellent agreement for large stripes and slow fronts.  The assumption of \eqnref{eqn:sqrassume} is invalid for fast fronts, and the $X/L_\bot$ profile becomes asymmetric (depressed on the leading edge, and bulging on the trailing edge) and both profiles show an overall supression of the chemical potential from the analytical prediction.  These additional profiles are not shown here.

However, to predict the values of $L$ and $U$ where nucleation occurs we only need the extremum value of the chemical potential.  This occurs in the center of a stripe at the front and is given by
\begin{align}
	& \mu(\vec{R}=\vec{0})
	= \pi U L_\bot \sum\limits_{n=-\infty}^{\infty}
		\left[ \frac{1}{2}\ln\left( \frac{16n^2 - 1}{16n^2 - 4} \right) \right. \\
		\nonumber
		& \left. {} + n\ln\left( \frac{32 n^3-6 n-1 }{32 n^3-6 n+1} \right) \right]
		= -2 U L_\bot \mathcal{C} = \mu^\text{peak}\;.
\end{align}
Here $\mathcal{C}=0.9159\ldots$ is a constant known as Catalan's constant.  Nucleation of a new stripe will occur when the chemical potential peak reaches the nucleation chemical potential $\mu^\text{peak}=\mu^\text{nucl}$.

Due to the symmetry at the center of the stripe, the nucleation chemical potential for a two-dimensional stripe is the same as the one-dimensional switching condition.  In a previous paper\cite{foard-2009} we presented an analytical expression for the switching concentration for the special case $M=0$ where there is negligible diffusive mobility ahead of the front.  In that case an analytical expression for the switching chemical potential can be found.  However, a nonzero $M$ induces earlier switching due to the presence of a nucleation kernel ahead of the front\cite{foard-2009}.  With the same method we used to verify the switching condition in our previous paper, we have measured the one-dimensional switching chemical potential $\mu^\text{nucl}\approx0.24$ for the non-dimensional parameters used in this two-dimensional system.  Thus we find that the maximum orthogonal stripe wavelength as a function of front speed is:
\begin{equation}
	L_{\bot}^\textrm{max} \approx \frac{0.24}{2\mathcal{C}U} \approx \frac{0.131}{U}\;.
\end{equation}

It is interesting to note that the analytical composition profile ahead of a solidification front for lamellar morphologies of eutectic mixtures presented by Jackson and Hunt is also a solution to the full chemical potential profile, although in a very different form\cite[Eqn.\ (3)]{jackson-1966}.  By using the substitutions $C \rightarrow \mu$, $S_\alpha = S_\beta \rightarrow L_\bot/4$, $C_0^\alpha = C_0^\beta \rightarrow 1/2$, $v \rightarrow U$, $d \rightarrow 1/ 2 \pi^2$, $x \rightarrow Y$, and $z \rightarrow \abs{X}$, their solution can be written:
\begin{equation}
	\mu(L_\bot \vec{R}) = 2 U L_\bot
	\sum\limits_{n=1}^\infty
	\frac{1}{n^2}
	\sin \left( \frac{n \pi}{2} \right)
	\cos \left( 2n\pi Y \right)
	e^{- 2n\pi \abs{X}}
	\;,
\end{equation}
which is, at least in a preliminary numerical evaluation, equivalent to our solution, although we have not yet shown this analytically.  For off-critical situations a complication occurs: the inflow material induces a non-neutral wetting condition for the orthogonal stripes.  This can be clearly seen in \figref{fig:depletion}.  This situation requires additional considerations that will be discussed elsewhere.

\subsection{Minimum Orthogonal Stripe Size}
\begin{figure}
	\begin{center}
		\psfrag{L}{$L$}
		\psfrag{1}{$1$}
		\psfrag{10}{$10$}
		\psfrag{T}{$T$}
		\psfrag{1}{$1$}
		\psfrag{10}{$10$}
		\psfrag{100}{$100$}
		\psfrag{1000}{$1000$}
		\psfrag{10000}{$10000$}
		\psfrag{L(T,k=0.5)}{$L_\text{C}(T)$, $\kappa=0.5$}
		\psfrag{L(T,k=1.0)}{$L_\text{C}(T)$, $\kappa=1.0$}
		\psfrag{L(T,k=2.0)}{$L_\text{C}(T)$, $\kappa=2.0$}
		\psfrag{lsp}{$\lambda_{sp}$}
		\psfrag{L = 0.555T^1/3}{$L_\text{C}=0.555T^{1/3}$}
		\includegraphics[clip=true, width=\linewidth]{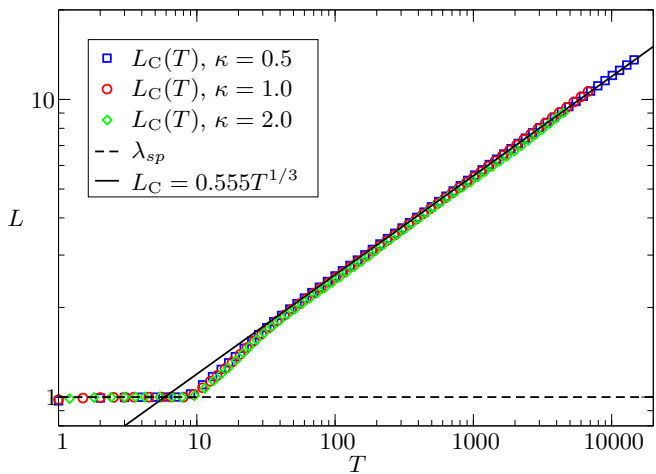}
		\caption{\label{fig:h_scaling}(Color online) Measurement of the dynamical scaling constant $\text{C}$ in \eqnref{eqn:bray} using LBM simulations.  Simulation size is $4096^2$ lattice sites except for $\kappa=2.0$ which is $1024^2$. The numerical method is similar to that which is outlined in \scnref{scn:simulation} for $u_x=0$, except with fully periodic boundaries and no front.  Characteristic length scale was measured by dividing the system area by the length of the $\Phi=0$ interface, then non-dimensionalized by multiplication with a scaling factor ($2.27/\lambda_{sp}$) such that $L_\text{C}=1$ is the length scale of spinodal decomposition.}
	\end{center}
\end{figure}

The minimum stripe width is limited by the width of stripes which, if a defect occurs, coarsen more quickly towards the front than the front moves away.  A simple qualitative argument for the defect speed can be obtained from the dynamical scaling laws\cite{bray-1994-43}.  This law describes the time evolution of a morphology by stating that at later times the structure is statistically similar to that of earlier times when scaled with the typical length scale
\begin{equation}\label{eqn:bray}
	L_\text{C} = \text{C} T^{1/3} \;.
\end{equation}
The constant $\text{C}$ is expected to depend on the parameters of the system and the details of the morphology.  From \eqnref{eqn:bray} we can define a coarsening speed which provides an estimate of how quickly the end of a finger of wavelength $L_\text{C}$ will recede:
\begin{equation}\label{eqn:coarsening}
	U_\text{C} = \frac{d}{dT}L_\text{C} = \frac{1}{3}\text{C} T^{-2/3} = \frac{1}{3}\text{C}^3L_\text{C}^{-2}\;.
\end{equation}
If a finger protrudes from the front into the phase separating region, it will coarsen in the same direction as the front.  If the coarsening speed is faster than the front $U_\text{C}>U$, the finger will eventually coarsen away completely into the mixed material domain.  According to \eqnref{eqn:coarsening} smaller fingers coarsen at a greater speed, and this sets a minimum wavelength for orthogonal stripes generated by a moving front:
\begin{equation}
	L_{\bot}^\textrm{min} \ge L_\text{C}(U) = 2 \sqrt{\frac{\text{C}^3}{3U}}\;.
\end{equation}
Unfortunately, numerical values of $\text{C}$ for different situations are hard to come by in the literature.  However, it is easy to obtain $\text{C}$ by measurement during phase ordering of a homogeneous quench.  We did this for a symmetric system (i.e. $\Phi(T=0)\approx0$) and obtained a value of $\text{C}\approx0.555$; for details please refer to \figref{fig:h_scaling}.

\begin{figure}
	\begin{center}
		\includegraphics[clip=true, width=\linewidth]{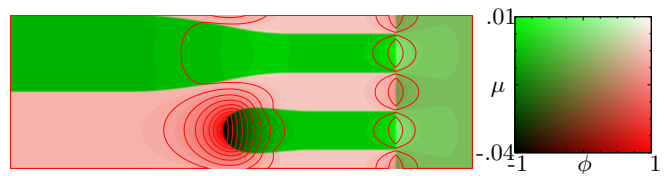}
		\caption{\label{fig:proto-stripe}(Color online) An example of a proto-stripe stabilized by regulation of front speed.  For the color version of this plot the range of values are shaded -- in the additive RGB color model -- red=$(\mu-\mu_{min})/(\mu_{max}-\mu_{min})$, green=$(\phi-\phi_{min})/(\phi_{max}-\phi_{min})$, blue=red$\times$green.  A  legend is shown at the right of the figure.  Equi-chemical potential lines are superimposed.  The front speed is increased when the stripe shrinks and decreased when the stripe grows until a stationary system is found; see text for methodology.  Initial conditions are similar to the final state (shown), with an initial front speed $U=(0.2L_{\bot})^2$ although the final $U$ was found to be insensitive to the initial value.  The result of several such simulations with varying initial $L_{\bot}$ are shown in \figref{fig:finger_coarsening}.  Each simulation used $\kappa = \pi^2/8\approx1.23$ (so that $u_x=U$) as the interfacial free energy parameter, though system size varies as a function of $L_{\bot}$. The parameters for this simulation were $l_x=312$, $l_y=104$ ($L=5.2687$), and $x_f=260$.  The final front speed was $U=0.00113$ giving a scaling constant $C=0.17746$.}
	\end{center}
\end{figure}

\begin{figure}
	\begin{center}
		\psfrag{L}{$L_\bot$}
		\psfrag{1}{$1$}
		\psfrag{10}{$10$}
		\psfrag{U}{$U$}
		\psfrag{0.001}{$0.001$}
		\psfrag{0.01}{$0.01$}
		\psfrag{0.1}{$0.1$}
		\psfrag{Simulation Result}{Simulation Result}
		\psfrag{L=0.18U^-1/2}{$L_\bot=0.18U^{-1/2}$}
		\includegraphics[clip=true, width=\linewidth]{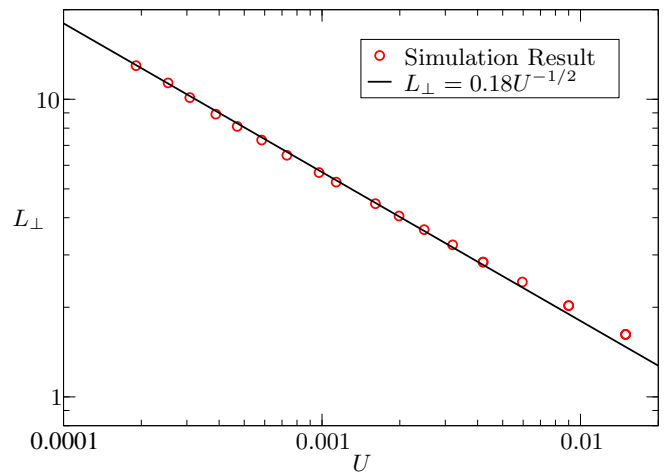}
		\caption{\label{fig:finger_coarsening}(color online) Measurement of the scaling constant $\text{C}$ for the finger morphology shown in \figref{fig:proto-stripe}.  A proto-stripe morphology of wavelength  $L_\bot$ is stabilized in a LBM simulation by adjusting the front speed $U$ until a stationary profile is achieved.  The fitted line gives an estimate of the minimum orthogonal stripe wavelength $L_\bot^\text{min}$ for a given front speed $U$.}
	\end{center}
\end{figure}

We expect this coarsening speed to be on the order of, but not exactly equal to, the speed with which a single finger coarsens.  To obtain a better estimate of the coarsening speed of a finger we can measure the speed in a simulation.  To do that we set up a proto-stripe similar to what is shown in \figref{fig:proto-stripe}.
We then measure the position of the tip of the finger as the first zero-crossing of $\phi$ at the original $y$ position of the center of the finger.  We then vary the front speed slowly (once every $4l_x$ iterations) to stabilize the position of the finger tip.  Once the velocity has reached a stationary state (determined by the tip speed and the average of the last 100 tip speed measurements being less than $U/10^6$) we find the coarsening speed of the finger.  The results of those measurements are shown in \figref{fig:finger_coarsening} and, as expected, the stripe speed is well approximated by  $U = 4 \text{C}^3 / 3 L_{\bot}^2$, with $\text{C}=0.29$.
This is approximately a factor of $2$ smaller than the dynamical scaling constant found in \figref{fig:h_scaling}.

\subsection{Region of Stability}\label{scn:stability}
We now have a theoretical prediction for the minimum and maximum orthogonal stripe wavelength as a function of front speed.  These boundaries describe region of stability for orthogonal stripes.  Fronts moving faster than the speed at which the minimum and maximum stripes intersect $U_\bot^\text{max}=0.52$ will not form orthogonal stripes, although orthogonal stripe may be formed by other influences after the front passes: for example the presence of external walls, etc. as in \cite{gonnella-2010}.

\begin{figure}
	\begin{center}
		\subfigure[Initial configuration]{\label{fig:hd-init}\includegraphics[clip=true, width=0.48\linewidth]{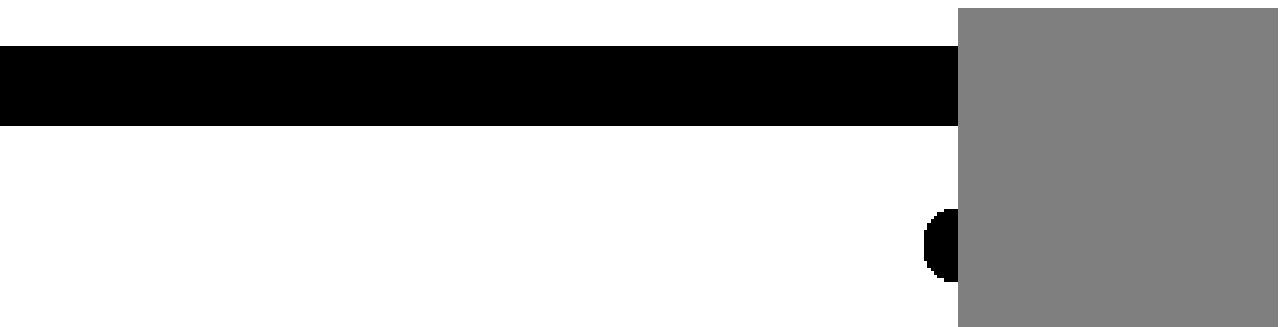}}
		\hfill
		\subfigure[Stable stripe]{\label{fig:hd-stable}\includegraphics[clip=true, width=0.48\linewidth]{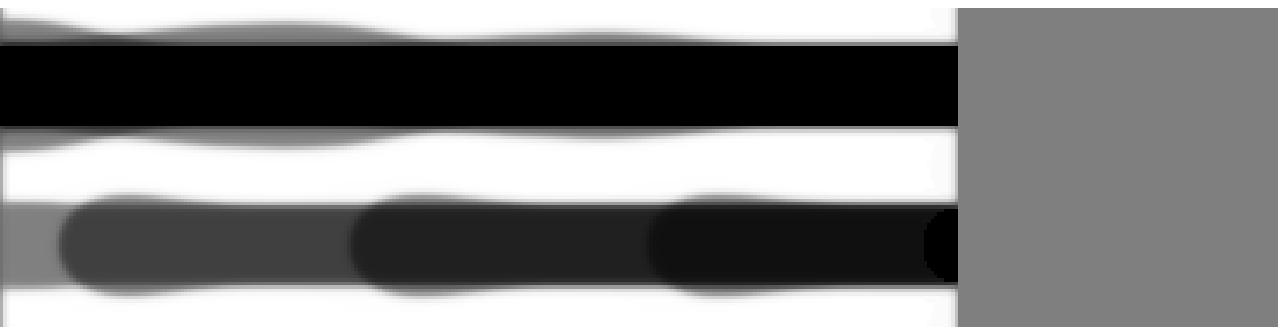}}

		\subfigure[Merged via coarsening]{\label{fig:hd-merge}\includegraphics[clip=true, width=0.48\linewidth]{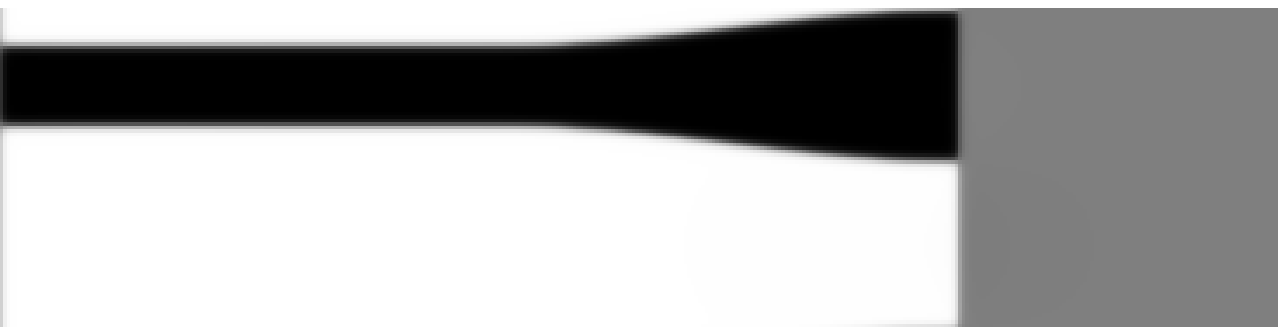}}
		\hfill
		\subfigure[Split via nucleation]{\label{fig:hd-split}\includegraphics[clip=true, width=0.48\linewidth]{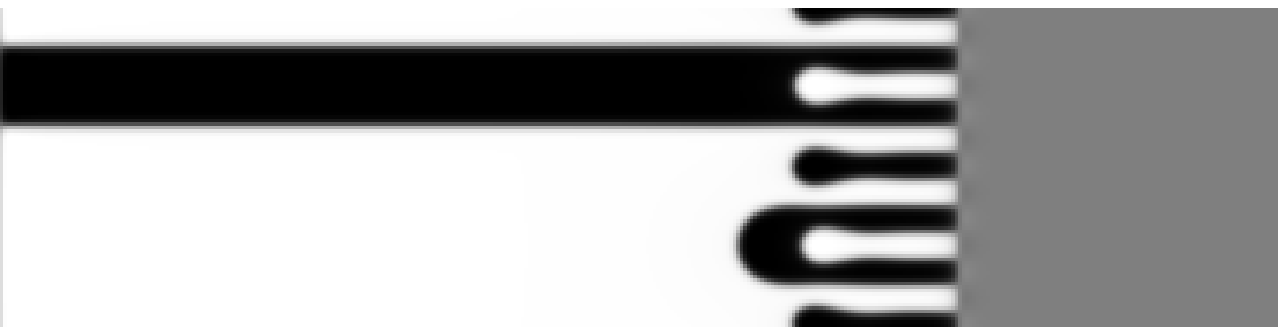}}
		\caption{\label{fig:hd}Examples of simulations performed to determine the stability of orthogonal stripes of a given size formed by a front moving at a prescribed speed in order to map the region of stable stripe formation in \figref{fig:orthogonal_theory}.  The initial configuration (a) is similar to the one used to predict the lower bound of the stability region (see \figref{fig:proto-stripe}), and those observed in Figs. \ref{fig:slow}, \ref{fig:depletion}, and \ref{fig:stripe}. The simulation is run until the number of $\mathcal{A}$-to-$\mathcal{B}$-type interfaces intersecting the front and the outflow boundary changes, then the simulation is halted and classified.  If the number of interfaces at the front decreases, the simulation is classified as having merged (c) and the point in \figref{fig:orthogonal_theory} gets a $\vartriangle$ symbol.  If this number increases the simulation is classified as split (d) and gets a $\triangledown$ symbol.  If the number of interfaces at the outflow boundary becomes 4 the simulation is classified stable (b) and the symbol is a circle with the radius proportional to the extremum value of the chemical potential at the front.}
	\end{center}
\end{figure}

\begin{figure}
	\begin{center}
		\psfrag{U}{$U$}
		\psfrag{0.001}{$0.001$}
		\psfrag{0.01}{$0.01$}
		\psfrag{0.1}{$0.1$}
		\psfrag{1}{$1$}
		\psfrag{L}{$L_\bot$}
		\psfrag{10}{$10$}
		\psfrag{Stable : r = m}{Stable: $r\propto\mu$}
		\psfrag{Too Small : Merged}{Merged: $L_\bot < L_\bot^\text{min}$}
		\psfrag{Too Large : Split}{Split: $L_\bot > L_\bot^\text{max}$}
		\psfrag{Lmax = 0.13U^-1}{$L_\bot^\text{max} = 0.13 U^{-1}$}
		\psfrag{Lmin = 0.18U^-1/2}{$L_\bot^\text{min} = 0.18U^{-1/2}$}
		\includegraphics[clip=true, width=\linewidth]{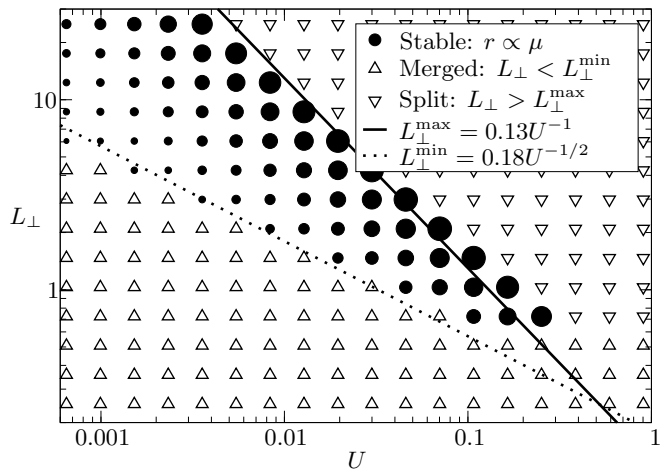}
		\caption{\label{fig:orthogonal_theory}Verification of minimum and maximum orthogonal stripe widths as a function of front speed for $\phi_{in}=0$.  Filled circles are at points where simulations demonstrated stable stripe formation.  Triangles are at points where stripe formation was unstable in the simulation.  See \figref{fig:hd} for a description of how the simulation results were obtained.  The predicted stable region is above the minimum (dashed) and below the maximum (solid) lines, and corresponds well with the field of filled circles.  See text for further discussion.}
	\end{center}
\end{figure}

We observe that the orthogonal stripes formed by a front in the simulation results shown so far fall in the region of stability, however we can test the bounds of the region more systematically.  We do this by testing points in the $L_\bot$ vs. $U$ parameter space for the formation of a stable stripe from the initial proto-stripe configuration shown in \figref{fig:hd-init}.
If the point is inside the region of stability, the proto-stripe grows as shown in the time-lapse overlay image in \figref{fig:hd-stable}.  If the point is below or above the region, the proto-stripe will respectively merge or split; examples of which are shown in (c) and (d) of \figref{fig:hd}.  If the point is to the right of the region, splitting and merging is followed by addition morphology changes until the stripes can reorient to become parallel to the front, however the simulation is classified and halted at the first morphology change.  The final results of this series of simulations is shown in \figref{fig:orthogonal_theory}.
These simulations confirm the predicted region of stability for orthogonal stripes for symmetrically mixed ($\Phi_{in}=0$) initial concentration.  Since our analytical predictions did not take into account the finite interface width we expect there will be deviation at small stripe widths and fast fronts.  We see that the prediction for $U_\bot^\text{max}=0.52$ is close to the maximum front speed capable of forming orthogonal stripes observed as $U_\bot^\text{obv}\approx0.3$ in \figref{fig:orthogonal_theory} or $U_\bot^\text{obv}\approx0.2$ in \figref{fig:surveymap-selective}.  The predicted allowed stripes widths for slow moving fronts agrees very well with simulation results.  However, a careful observer will notice that there is a slight discrepancy and our proto-stripe seem to allow for slightly smaller front speeds.  We attribute this detail to the shape of the persistent stripe, which is different for both kinds of simulations (see Figs. \ref{fig:proto-stripe} \& \ref{fig:hd}).  The curvature in \figref{fig:proto-stripe} will allow for a slightly faster coarsening of the stripe tip.

We have now determined the characteristics of the ordered two-dimensional morphologies observed to be formed by phase separation fronts moving into mixtures of critical concentration: the orthogonal stripe morphology just presented, and the parallel stripe morphology--an essentially one-dimensional structure which we described and analyzed in previous work\cite{foard-2009, foard-2010}.  The other ordered morphology we observed was an ordered droplet structure which was not observed to occur for fronts moving into critical mixtures.

\section{Outlook}\label{scn:outlook}
In this paper we presented a survey of the morphologies formed in the wake of a sharp phase-separation front. The resulting morphologies could be characterized as lamella in a parallel orientation with respect to the front, lamella in an orthogonal orientation and droplet arrays. We found that the selected morphology depended on the front speed, the volume fraction of the overtaken material, but also on the history of the system. If the front emerges from a homogeneous quench a depletion layer is typically formed and this will lead to the preferred formation of lamella oriented parallel to the front. We saw, however, that sometimes defects will form and two systems under the same conditions--where the only difference lies in the random initial conditions for the homogeneous phase-separation--can instead form orthogonal lamella. By providing an unbiased initial condition we were able to determine the ``preferred'', i.e. most stable morphology. Using these results we were then able to present a state diagram as a function of the front speed and the volume fraction.  We then examined in detail the formation of the orthogonal lamella for fronts moving into critical mixtures.  We determined the range of allowed lamella sizes for a given front speed by analytically predicting the minimum and maximum lamella which can be stably formed.  This gave a prediction for the transition between orthogonal and parallel lamella for critical mixtures which was within a factor of 3 of the observed transition point.

The next step in this analysis will be to determine the allowed stripe widths for fronts moving into mixtures with a minority and a majority phase in an effort to predict the boundary of the ``orthogonal stripes'' region in \figref{fig:surveymap-morphology}--now marked with an observed dashed line.  This will require additional research, such as:  Determining how dynamical coarsening of stripe morphologies is changed by having off-critical mixtures; this subject is not well studied as dynamical scaling is typically studied in the context of homogeneous quench, which for off-critical mixtures results in droplet morphologies that coarsen due to Ostwald ripening\cite{bray-1994-43}.  Determining what effect off-critical mixtures ahead of the front will have on the nucleation of new stripe domains.  Determining how the stripe morphology itself is altered by having an off-critical mixture.  This last point may seem trivial at first glance, as conservation of the order parameter requires for stable stripes to form, the final width of the minority and majority stripes are a simple function of the mixed material concentration.  We observe this to be true in our simulations for some distance away from the front, but this is not the case directly at the front where the dynamics of morphology selection occur.  As shown most readily in \figref{fig:depletion}, but also elsewhere in this paper, there is a pinching off of the minority phase due to the presence of a preferred contact angle induced by the control parameter front.  These considerations will be published in future work.

This paper raises a number of interesting issues. Firstly, can one predict \emph{a priori}, which morphology is most stable, and thereby provide analytical predictions for the state diagram? Secondly, what are the limits of metastability, i.e. what is the fastest orthogonal lamella structure that can be formed and what is the slowest parallel lamella morphology that can be formed and how far can either encroach on the droplet states and vice versa?  Thirdly, what is the region of stability for orthogonal lamella formed in off-critical mixtures, and by determining this region can we predict the boundary between orthogonal lemella droplet morphologies?  Additionally, it would be very useful to more accurately determine where in the parameter space of this, and perhaps other similar models, is the transition from parallel to orthogonal stripe morphologies.  This transition has been noted several times, but a systematic study has not been done.

So far we have been able to successfully analytically predict the size of parallel lamella structures. In future work we will refine our prediction for the extent of the orthogonal stripe region in the morphology diagram by considering the effect of the non-neutral wetting condition at the front for off-critical volume fractions.  Another natural extension is to consider the system in three dimensions. In a future paper we will present an analogous study which shows a  slightly richer state diagram which includes cylinder arrays as well as three dimensional droplet lattices. Lastly this paper only considered diffusive dynamics. For many practical application it is important to include hydrodynamics effects, which can alter domain formation considerably.

\begin{acknowledgments}
E. M. Foard wishes to graciously thank Giuseppe Gonnella for many insightful comments and helpful discussions, and Goetz Kaehler for constructive advice during the preparation of this manuscript.  This research is funded in part by a ND EPSCoR seed grant.  
\end{acknowledgments}

\bibliography{Foard_Wagner_Morphologies-in-2D}

\end{document}